%% file: main.tex
\documentclass{article}

\usepackage[table]{xcolor} % Enable table coloring
\usepackage{graphicx} 
\usepackage{times}
\usepackage{tikz}
\usepackage{makecell}
\usepackage{listings}
\usepackage{tcolorbox}
\usepackage{algorithm}
\usepackage{algpseudocode}
\usepackage{algorithm}
\usepackage{algorithmicx}
\usepackage{algpseudocode}
\usepackage{nicematrix}

\usepackage[final]{EMNLP2023}
\usepackage{amsmath}
\usepackage{amssymb}
\usepackage{mathtools}
\usepackage{amsthm}
\usepackage{enumitem}
\usepackage{subcaption}
\usepackage{wrapfig}

\usepackage{comment}
\usepackage[utf8]{inputenc} % allow utf-8 input
\usepackage[T1]{fontenc}    % use 8-bit T1 fonts
\usepackage{hyperref}       % hyperlinks
\usepackage{url}            % simple URL typesetting
\usepackage{booktabs}       % professional-quality tables
\usepackage{amsfonts}       % blackboard math symbols
\usepackage{nicefrac}       % compact symbols for 1/2, etc.
\usepackage{microtype}      % microtypography
\usepackage{xcolor}         % colors
\usepackage{multirow} 
\usepackage{graphicx}
\usepackage{booktabs}
\usepackage{tabularx}
\usepackage{pifont}
\definecolor{champagne}{RGB}{247, 231, 206} % Define champagne color

\usepackage{etoc}
\etocdepthtag.toc{mtchapter}
\etocsettagdepth{mtchapter}{subsection}
\etocsettagdepth{mtappendix}{none}

\definecolor{mblue}{RGB}{0, 61, 124}
\definecolor{myellow}{RGB}{239, 124, 0}
\definecolor{mnavy}{RGB}{0,0,128}
\definecolor{minc}{RGB}{0,128,0}
\definecolor{mdec}{RGB}{255,0,0}
\definecolor{mhold}{RGB}{128,128,128}
\definecolor{darksalmon}{rgb}{0.91, 0.59, 0.48}
\definecolor{emerald}{rgb}{0.31, 0.78, 0.47}
\definecolor{green(pigment)}{rgb}{0.0, 0.65, 0.31}
\definecolor{amaranth}{rgb}{0.9, 0.17, 0.31}
\definecolor{iris}{rgb}{0.35, 0.31, 0.81}
\definecolor{uu}{rgb}{0.95, 0.51, 0.51}
\definecolor{spirodiscoball}{rgb}{0.06, 0.75, 0.99}

\newcommand{\equal}{Equal Contribution}

\title{MegaAgent: A Large-Scale Autonomous LLM-based Multi-Agent System Without Predefined SOPs}
\author{
Qian Wang\textsuperscript{1}\thanks{\equal}, 
Tianyu Wang\textsuperscript{2}\footnotemark[1], 
Zhenheng Tang\textsuperscript{3}, \\
\textbf{Qinbin Li\textsuperscript{4},} 
\textbf{Nuo Chen\textsuperscript{1},} 
\textbf{Jingsheng Liang\textsuperscript{5},} 
\textbf{Bingsheng He\textsuperscript{1}}\\
\textsuperscript{1}National University of Singapore \quad \textsuperscript{2}Zhiyuan College, Shanghai Jiao Tong University \\
\textsuperscript{3}The Hong Kong University of Science and Technology \\
\textsuperscript{4}Huazhong University of Science and Technology \\
\textsuperscript{5}South China University of Technology \\
}

\begin{document}

\maketitle

\begin{abstract}
LLM-based multi-agent systems (MAS) have shown promise in tackling complex tasks. However, existing solutions often suffer from limited agent coordination and heavy reliance on predefined Standard Operating Procedures (SOPs), which demand extensive human input. To address these limitations, we propose \textit{MegaAgent}, a large-scale autonomous LLM-based multi-agent system. \textit{MegaAgent} generates agents based on task complexity and enables dynamic task decomposition, parallel execution, efficient communication, and comprehensive system monitoring of agents. In evaluations, \textit{MegaAgent} demonstrates exceptional performance, successfully developing a Gobang game within 800 seconds and scaling up to 590 agents in a national policy simulation to generate multi-domain policies. It significantly outperforms existing systems, such as MetaGPT, in both task completion efficiency and scalability. By eliminating the need for predefined SOPs, \textit{MegaAgent} demonstrates exceptional scalability and autonomy, setting a foundation for advancing true autonomy in MAS.\footnote{Code is available at \url{https://github.com/Xtra-Computing/MegaAgent}.}
\end{abstract}

\input{sec-introduction}

\input{sec-methodology}

\input{sec-experiments}
\input{sec-relatedwork}
\input{sec-conclusion}

% \bibliographystyle{ACM-Reference-Format}
% \bibliography{bibliography}

\bibliographystyle{acl_natbib}
\bibliography{bibliography}

%%%%%%%%%%%%%%%%%%%%%%%%%%%%%%%%%%%%%%%%%%%%%%%%%%%%%%%%%%%%

%%%%%%%%%%%%%%%%%%%%%%%%%%%%%%%%%%%%%%%%%%%%%%%%%%%%%%%%%%%%
\clearpage
\appendix
\input{sec-appendix}

\end{document}

%% file: sec-introduction.tex
\section{Introduction} \label{intro}
The remarkable planning and cognitive capabilities of Large Language Models (LLMs) \cite{touvron2023llamamodel, zhu2023minigpt} have spurred significant interest in LLM-based multi-agent systems (MAS) \cite{wu2023autogen, chen2023agentverse, hong2023metagpt}, which coordinate multiple LLM agents to address complex tasks. For example, MetaGPT introduces a meta-programming framework to simulate the software development process \cite{hong2023metagpt}, while Simulacra \cite{park2023generative} models social interactions among 25 LLM-powered agents in a simulated town, showcasing the potential of these systems to replicate real-world dynamics. The demand for large-scale social simulation applications, such as social media and war simulations \cite{gao2023s, hua2023war, jin2024if}, is driving the development of MAS capable of simulating complex real-world scenarios.

However, existing MAS have two limitations. (1) They fail to achieve adaptive task coordination when the task is big and complex e.g. generating hundreds of agents for a social simulation; and do not consider the coordination between a large scale of agents.
(2) Most systems heavily depend on user-defined configurations, including predefined agent roles, standard operating procedures (SOPs), and static communication graphs \cite{hong2023metagpt, chen2023agentverse, wu2023autogen}. This approach limits flexibility and requires significant human effort when deploying numerous agents to complete a task. 

Addressing the above limitations presents the following key challenges: (1) \textbf{Facilitating adaptive and effective communication among agents and with external file systems.} As tasks grow in complexity and scale, managing communication becomes increasingly difficult, especially when incorporating parallelism and coordinating multiple agents across different rounds of communication \cite{zhang2024towards}. (2) \textbf{Ensuring that each agent completes its task accurately without relying on predefined SOPs.} LLM agents often generate hallucinated outputs \cite{huang2023survey} or fail to complete tasks correctly within a single round \cite{liu2023agentbench, andriushchenko2024agentharm}, necessitating robust mechanisms to ensure reliability and correctness. This is particularly critical in multi-agent systems, where hallucinations can propagate and compromise the entire system's performance \cite{zhang2024towards, lee2024prompt}.

\begin{figure*}
    \centering
\includegraphics[width=0.85\textwidth]{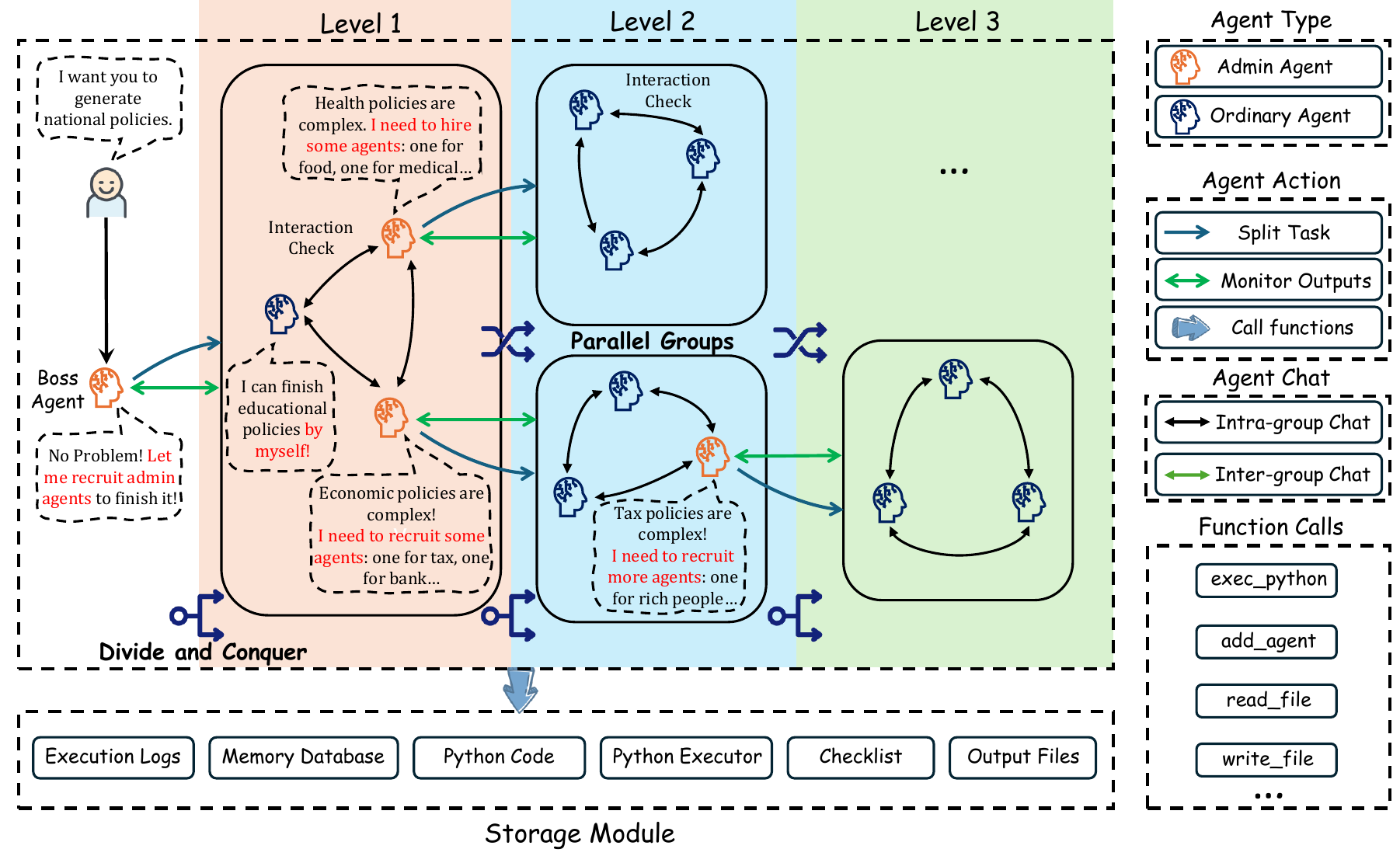}
    % \vspace{-0.5em}
    \caption{MegaAgent processes a user-provided meta-prompt by dividing it into distinct tasks, assigning each to a corresponding admin agent. Admin agents oversee their tasks, autonomously recruiting additional agents as needed to form task-specific groups that operate in parallel for efficient execution. These groups can further expand through sub-agent recruitment, creating a multi-level hierarchy. Admin agents supervise their groups to ensure task completion and output quality. Agents are classified into admin and ordinary agents: admin agents can communicate with one another, while ordinary agents interact only within their groups to optimize communication efficiency. Each agent is treated as a process whose state is either \emph{Idle}, \emph{Processing}, or \emph{Response}. Agents access and manage external files in storage module using function calls, supporting seamless data retrieval and task execution.}
    \label{fig:overview}
    % \vspace{-1em}
\end{figure*}

Drawing inspiration from Operating Systems (OS), where processes and threads efficiently manage tasks through: (1) generating multiple threads within a process to complete a task, (2) enabling different processes to operate in parallel, and (3) operating a producer-consumer-based message queue, we propose \textit{MegaAgent} to address the aforementioned limitations. \textit{MegaAgent} decomposes large tasks into multiple hierarchical subtasks (analogous to processes), with each subtask completed by a dedicated group of agents (similar to threads). Communication occurs either within agent groups or between them as needed, resembling inter-process communication in an OS. \textbf{Users simply need to provide a meta prompt to  \textit{Boss Agent}, after which the task is autonomously completed.} The novelty comparison between \textit{MegaAgent} and popular baselines is in \autoref{tab:relate-work-simple}. Details are in \autoref{tab:relate-work}. An overview of \textit{MegaAgent} is shown in \autoref{fig:overview}. We equip \textit{MegaAgent} with the following two strategies to tackle the above challenges: 

(1) \textbf{Hierarchical Task Management:} 
To facilitate adaptive task handling and effective communication, \textit{MegaAgent} employs a hierarchical task management mechanism structured across three levels. First, the \textit{Boss Agent Level Task Decomposition}, where the Boss Agent divides a task into smaller subtasks and assigns them to admin agents. Next, \textit{Dynamic Hierarchical Group Formation} occurs when an admin agent recruits additional agents if a subtask exceeds its capacity, forming a recursive, dynamic group to handle complex tasks. Finally, \textit{System-Level Coordination and Communication} enables parallel execution and efficient interaction across the system, with agents connecting to external systems through function calls.

(2) \textbf{Hierarchical Monitoring:}  
To ensure agents complete tasks accurately without relying on predefined SOPs, \textit{MegaAgent} incorporates hierarchical monitoring and coordination mechanisms. First, each agent is assigned a task by its admin agent upon generation. Then, \textit{MegaAgent} employs a robust hierarchical monitoring and coordination framework for each agent as follows: \textit{Agent-Level Monitoring} ensures each agent tracks its actions with a checklist, verifying progress before moving forward. At the \textit{Group-Level Monitoring}, admin agents oversee the tasks of their assigned agents, ensuring smooth execution and coordination. At the \textit{System-Level Monitoring}, the Boss Agent reviews the outputs of all groups to ensure accuracy and consistency. This hierarchical approach ensures both effective task management and completion across all levels.

\begin{table}[!ht]
\centering
\resizebox{\linewidth}{!}{%
\begin{tabular}{c|cccc}
\toprule
\rowcolor{champagne}\textbf{Model} & \textbf{No Predefined SOP} & \textbf{Multi-file Support} & \textbf{Parallelism} & \textbf{Scalability} \\
\midrule
\rowcolor{gray!10}AutoGen  & \textcolor{darksalmon}{\ding{55}} & \textcolor{darksalmon}{\ding{55}} & \textcolor{darksalmon}{\ding{55}} & \textcolor{darksalmon}{\ding{55}}\\
MetaGPT  &  \textcolor{darksalmon}{\ding{55}} &\textcolor{green(pigment)} {\ding{51}} & \textcolor{darksalmon}{\ding{55}} & \textcolor{darksalmon}{\ding{55}}\\
\rowcolor{gray!10}CAMEL & \textcolor{darksalmon}{\ding{55}} & \textcolor{darksalmon}{\ding{55}} & \textcolor{darksalmon}{\ding{55}}  & \textcolor{darksalmon}{\ding{55}} \\
AgentVerse  & \textcolor{green(pigment)} {\ding{51}} & \textcolor{darksalmon}{\ding{55}} & \textcolor{darksalmon}{\ding{55}}  & \textcolor{darksalmon}{\ding{55}} \\
\midrule
\rowcolor{gray!10}\textbf{MegaAgent} & \textcolor{green(pigment)} {\ding{51}} & \textcolor{green(pigment)} {\ding{51}} & \textcolor{green(pigment)} {\ding{51}}  & \textcolor{green(pigment)} {\ding{51}}\\
\bottomrule
\end{tabular}
}
\caption{Novelty comparison of popular MAS with MegaAgent. Details are explained in \autoref{tab:relate-work}.}
\label{tab:relate-work-simple}
\end{table}

MegaAgent's framework is most beneficial for future applications requiring large-scale coordination of autonomous agents. In financial markets, it models systems where agents represent traders, investors, or regulatory bodies. These agents process market news, analyze trends, make trades, and respond to macroeconomic events, potentially scaling to millions or billions of agents. Initial studies highlight its applicability \cite{zhang2024ai, gao2024simulating}. In healthcare, MegaAgent can optimize systems with agents representing providers, patients, administrators, and policymakers, simulating complex workflows \cite{li2024agent, fan2025ai}.

We demonstrate \textit{MegaAgent}'s effectiveness and autonomy through comprehensive experimental validation. First, we establish its strong foundational capabilities by evaluating its performance against leading baselines on several standard benchmarks: MBPP \cite{austin2021program}, HumanEval \cite{chen2021evaluating}, MATH \cite{hendrycks2021measuring}, and GSM-8K \cite{cobbe2021training}. Subsequently, to showcase its advanced coordination abilities, we conduct two experiments in widely recognized MAS research scenarios demanding complex multi-agent coordination \cite{hong2023metagpt, guo2024large}: \textbf{(1) Software development: Gobang Game Development.} This experiment highlights \textit{MegaAgent}'s superior autonomy and efficiency, being \textbf{the only model capable of completing the task within 800 seconds.} \textbf{(2) Social Simulation: National Policy Generation.} This task demonstrates \textit{MegaAgent}'s large-scale autonomy and scalability, \textbf{generating and coordinating approximately 590 agents to produce the expected policies within 3000 seconds.} In contrast, baseline models struggle with these complex tasks, often coordinating fewer than 10 agents and failing to generate the expected outcomes.

Our contributions are as follows:

\begin{itemize}
    \item[\ding{182}] \textbf{\textit{Autonomous Framework.}} We introduce \textit{MegaAgent}, a large-scale autonomous LLM-based multi-agent system. It supports dynamic task decomposition, parallel execution, and systematic monitoring, ensuring efficient task management.
    \item[\ding{183}] \textbf{\textit{Minimizing Human-designed Prompts.}} We notice the importance of minimizing human-designed prompts in MAS, addressing a critical limitation of previous frameworks that creates a bottleneck for large-scale MAS for complex tasks. To overcome this, we propose assigning LLM agents to autonomously split tasks and generate SOPs for agents. This approach reduces human intervention and enable broader range of users to employ MAS effectively.
    \item[\ding{184}] \textbf{\textit{Comprehensive Experimental Validation.}}  
    Extensive experiments demonstrate \textit{MegaAgent}'s capabilities:  
    \textbf{(1) Strong Foundational Performance}: \textit{MegaAgent} consistently outperforms leading baselines on standard benchmarks such as MBPP, HumanEval, MATH, and GSM-8K, showcasing its robust problem-solving abilities.
    \textbf{(2) Superiority in Complex Coordination}: In demanding multi-agent scenarios, it is the only framework capable of completing both Gobang game development and national policy simulation tasks, outperforming all baselines.  
    \textbf{(3) Efficiency and Scalability}: \textit{MegaAgent} successfully completes the Gobang game development task within 800 seconds. Furthermore, it efficiently coordinates approximately 590 agents for national policy generation within 3000 seconds, while baselines manage fewer than 10 agents and fail to complete the task. This remarkable agent count underscores \textit{MegaAgent}'s scalability.
\end{itemize}

% \begin{itemize}
%     \item We highlight the importance of minimizing human-designed prompts in MAS, addressing a critical limitation of previous frameworks that creates a bottleneck for users from non-computer science domains by assigning LLM agents to split and asigning the SOPs for the agents autonomously, with 
%     \item  We introduce MegaAgent, a practical framework for autonomous coordination in MAS. MegaAgent enables dynamic task decomposition and parallel execution, significantly improving efficiency and scalability.
%     \item We validate \ourmethod performance through experiments in software development and social simulations, demonstrating its superior autonomy, scalability, and adaptability compared to existing frameworks.
%     % \item \textbf{Future Directions}: We provide insights into the challenges of scaling MAS systems and outline potential research directions to advance the field further.
% \end{itemize}

The remainder of this paper is organized as follows: Section~\ref{framework} introduces \textit{MegaAgent} framework in detail. Section~\ref{sec:experiments} presents experimental evaluations demonstrating \textit{MegaAgent}'s effectiveness. Section~\ref{sec:related-work} reviews related work, and Section~\ref{sec:conclusion} concludes the paper.

%% file: sec-methodology.tex
\section{MegaAgent Framework} \label{framework}

\subsection{Overview}
We introduce the \textit{MegaAgent} framework from two hierarchical perspectives, as outlined in Section~\ref{intro}: (1) Hierarchical Task Management and (2) Hierarchical Monitoring. An overview is provided in \autoref{fig:overview}. We elaborate the algorithm of MegaAgent in Algorithm~\ref{alg:megaagent_revised}.

% \textcolor{blue}{As illustrated in Figure \ref{fig:overview}, MegaAgent addresses the limitations of existing LLM-MA systems, such as rigid task definitions, lack of parallel execution, and scalability constraints, by introducing five key components: (1) Multi-Level Task Splitting, which processes the user's prompt and dynamically decomposes tasks into manageable subgroups, enabling flexible and efficient task allocation; (2) Action Module, composed of specialized agent groups, each focusing on a specific subtask to ensure parallel and autonomous execution; (3) Storage Module, integrating databases, files, and checklists to provide persistent memory and streamline task tracking; (4) Monitor Mechanism, which includes OS agents, admin agents, and checklists to oversee and coordinate agent actions, ensuring accountability and error management; and (5) Communication Mechanism, facilitating seamless interactions both between agents and with external services. These components work together to overcome the sequential bottlenecks and lack of scalability in existing approaches, enabling MegaAgent to achieve unprecedented levels of autonomy and efficiency. Each component is detailed below.}

\subsection{Hierachical Task Management}

\subsubsection{Multi-level Task Splitting}

To efficiently manage complex tasks in large-scale MAS, we implement a multi-level task management framework. \textit{Boss Agent} is responsible for decomposing the main task into manageable subtasks upon receiving the meta-prompt from a user. Each subtask is delegated to a specialized admin agent with a well-defined role by \textit{Boss Agent}. If a subtask is too complex for an admin agent to complete independently, it can recruit additional agents to handle specific components. These newly created agents can, in turn, recruit more agents if needed, assuming the role of admin agent themselves, as depicted in \textit{Level 2} and \textit{Level 3} in \autoref{fig:overview}. This recursive task-splitting mechanism enables the system to adapt dynamically as task complexity increases.

To enhance efficiency, we implement a parallel mechanism for agent groups operating at the same level. For instance, the two agent groups in \textit{Level 2} of \autoref{fig:overview} can work in parallel, with one generating economic policies and the other developing health policies. This parallelization reduces overall task completion time.

% \begin{figure}[htp]
% \centering
% \begin{tikzpicture}
%     \definecolor{shadecolor}{gray}{0.95}

%     % Define the node for MinisterHealth
%     \node[
%         align=left,
%         text width=0.45\textwidth,
%         fill=shadecolor,
%         rounded corners=1mm,
%         draw=black,
%         line width=0.5mm
%     ] (node1) at (0,0) {
%         \textit{You are MinisterHealth, the Health Policy Minister of the Health Department. Your job is to develop a comprehensive health policy document \texttt{`policy\_health.txt'} according to specified guidelines. You will collaborate with relevant departments (details omitted here).}
%     };

% \end{tikzpicture}
% \caption{Task Assignment for Health Policy Development.}
% \label{fig:health_policy_task}
% \end{figure}

\subsubsection{Hierarchical Coordination Mechanism}
Effective task execution in \textit{MegaAgent} is driven by a two-layer hierarchical coordination structure: (1) \textit{Intra-group Chat}, where agents within the same task group collaborate by sharing updates through prompt-based communication, ensuring smooth progress and effective task execution when interaction is required, as indicated by the black double-arrow line in \autoref{fig:overview}; and (2) \textit{Inter-group Chat}, where admin agents from different groups communicate to resolve task dependencies and coordinate cross-group efforts, as represented by the yellow double-arrow line in \autoref{fig:overview}. For instance, in the software development experiment discussed in \autoref{soft-exp}, the software implementation must adhere to the game logic designer's requirements. Ordinary agents are restricted from directly communicating with agents outside their group to enhance efficiency. % Each agent have a message queue, into which messages are sent using a typical Producer-Consumer pattern.

\subsubsection{Message Queue Mechanism} For each agent, we implement a message queue system based on the \textit{producer-consumer} paradigm to manage asynchronous communications. External agents act as \textit{producers} that generate messages through function calls, while the target agent acts as the \textit{consumer} that periodically retrieves batched messages via a polling mechanism. This architecture introduces three operational states: \textbf{1. Idle State}, where the consumer agent has no message, and will not have computational cost; \textbf{2. Processing State:} Upon message acquisition, the agent enters an atomic execution phase where LLM inference engines process the message batch. The message received during this state will stay in the message queue until the next processing state; \textbf{3. Response State:} Processed outputs undergo validity verification before being dispatched through designated function call. After that, the agent will check whether its message queue is empty. If so, it will enter the Idle state; if not, it will enter the Processing state. This approach can merge multiple messages to process at one time while sufficiently utilize the throughput of the LLM, thus balancing the cost and efficiency. 

\subsubsection{File Management}
To enable effective interaction between LLM agents and external files, we introduce an external storage module that manages all file-related tasks. This module includes components such as agent execution logs, a memory database, task monitoring tools, Python code execution support, shared files, and individualized agent checklists. To ensure consistent and accurate file management, we propose the following two designs:

(1) \textbf{Git-Based Version Control.}  To maintain file consistency, we integrate a Git-based version control mechanism. Since agents may spend considerable time editing files after reading them, concurrent modifications by other agents could cause conflicts. To prevent this, an agent retrieves the file's current Git commit hash upon reading it. Before making changes, the agent submits this hash to the file management system, which commits the updates, merges them into the latest HEAD, and prompts the agent to resolve any merge conflicts if necessary. All Git operations are serialized using a global mutex lock to ensure synchronization and prevent race conditions.

(2) \textbf{Long-Term Memory Management with a Vector Database.} Many studies show that LLM agents would forget the conversation history after several rounds due to the token length limit \cite{becker2024multi, xue2024repeat}. To address this, we implement a vector database to store the outputs of agents. Each output is encoded into embeddings using language models and stored in a vector database. Therefore, agents can retrieve relevant memory entries, enabling them to recall past interactions and maintain contextual awareness.

To visually illustrate the communication, we present an example of agent communication in \autoref{fig:communication}.

\begin{figure}[htp]
\centering
\begin{tikzpicture}
    \footnotesize
    \definecolor{chatcolor1}{HTML}{FFFACD} % Highlight color for the prompt
    \definecolor{shadecolor}{gray}{0.95} % Background color for context
    \fontfamily{cmss}\selectfont
    % Formatting for the user's prompt
    \node[align=left, text width=0.45\textwidth, fill=shadecolor, rounded corners=1mm, draw=black, line width=0.5mm, fill opacity=0.7, text opacity=1, anchor=north west] (node1) at (0,0) {
        \textit{You are MinisterHealth, the Minister of Health. Your job is to develop a comprehensive policy document ('policy\_health.txt') according to the guidelines provided in 'policy\_health.txt'. You will collaborate with MinisterEducation (the Minister of Education), MinisterFinance (the Minister of Finance), and pass the final document to MinisterInfrastructure (the Minister of Infrastructure). You can recruit lots of citizens for testing health initiatives. Ensure adherence to the specified routine only. Your collaborators include MinisterEducation, MinisterFinance, and MinisterInfrastructure.}
    };
    % Placeholder for generated responses or annotations    
\end{tikzpicture}
\vspace{-5pt}
\caption{Agent Communication Example: \textit{National Leader} to \textit{Minister of Health}}
\label{fig:communication}
\end{figure}

\subsection{Hierarchical Monitoring}
To ensure accurate task execution and minimize the propagation of hallucinations \cite{huang2023survey, hao2024quantifying} in an MAS, we implement a hierarchical monitoring mechanism that facilitates real-time oversight, error correction, and progress validation through a structured process.

\subsubsection{Multi-level Monitoring}

The monitoring system in \textit{MegaAgent} follows a structured, multi-level hierarchy to ensure accurate task completion and prevent error propagation. Then, \textit{MegaAgent} employs a hierarchical monitoring and coordination framework for each agent as follows:  
\begin{itemize}
    \item \textit{Agent-Level Monitoring:} Each agent maintains an \textit{checklist} upon its being generated by its admin agent to document its actions and verify progress. This monitoring ensures accountability and allows agents to independently validate their work before proceeding to the next step.
    \item \textit{Group-Level Monitoring:} Each agent group is supervised by an admin agent, which tracks the progress of individual agents, ensures smooth execution, and coordinates tasks within the group. 
    \item \textit{System-Level Monitoring:} At the highest level, \textit{Boss Agent} oversees the outputs of all agent groups upon task completion, ensuring adherence to the correct format and minimizing hallucinated results. This process enhances system-wide consistency, reliability, and correctness.
\end{itemize}

\subsubsection{Failure Scenarios and Solutions}

Monitoring focuses on two key aspects: output format verification and result validation, detailed as follows:

(1) \textbf{Output Format Verification.}  First, the monitoring would focus on the output format of an agent. For example, if an agent generates a Python file that fails to execute, its admin agent would flag the issue, log the error, and prompt a retry. By enforcing consistent output formats, this step prevents downstream agents from misinterpreting data, reducing potential hallucinations.

(2) \textbf{Result Validation.} Once a group completes its tasks, the admin agent reviews the generated outputs and compares them against the initial task requirements. If discrepancies are detected, the admin agent would detail error messages, outline missing or incorrect aspects, prompt the responsible agents to revise their work. This validation process ensures that final outputs align with intended objectives while minimizing task failures.  

To clarify the monitoring process, we outline common failure scenarios and solutions as follows:

\begin{itemize}  
    \item \textit{Incomplete TODO Lists:}  
    Agents may terminate prematurely or enter infinite loops due to inherent LLM limitations. An admin agent would detect it and prompt the agent to retry the task to ensure task completion.  

    \item \textit{Task Repetition:}  
    Limited context memory may cause agents to forget completed tasks, leading to redundant actions or task loops. An admin agent would identify inconsistencies by cross-referencing agent checklists and prompts corrective actions as necessary.  

    \item \textit{Secure Alignment Interruptions:}  Agents may become unresponsive or repeatedly return alignment-related constraint messages, such as \textit{"Sorry, I can't help with that."} In such cases, an admin agent attempts to recruit other agents to finish the task.
\end{itemize}  

To visually illustrate the monitoring, we present a monitoring log example in \autoref{fig:monitor}.

\begin{figure}[htp]
\centering
\begin{tikzpicture}
    \definecolor{chatcolor1}{HTML}{FFFACD}
    \definecolor{shadecolor}{gray}{0.95}

    % Define the first node
    \node[
        align=left,
        text width=0.45\textwidth,
        fill=shadecolor,
        rounded corners=1mm,
        draw=black,
        line width=0.5mm
    ] (node1) at (0,0) {
        \textit{**Financial Policy**
   - **Strengths**: Clear objectives, funding sources, and metrics for success.
   - **Improvements Needed**: 
     - Ensure curriculum relevance for different demographics.
     - Incorporate interactive learning elements.
     - Make programs accessible to all, including those with disabilities.
     - Establish a feedback loop for continuous improvement.
   - **Action**: Talk to MinisterFinance for revisions based on feedback.}
    };

\end{tikzpicture}
\caption{Monitoring Log Example: \textit{National Leader} monitors the financial policy.}
\label{fig:monitor}
\end{figure}

By combining strict output format verification and result validation, this monitoring framework ensures agents remain aligned with system goals. Comprehensive error-handling processes prevent cascading failures, ensuring system stability and optimal performance throughout the MAS.

%% file: sec-experiments.tex
\section{Experiments}
\label{sec:experiments}
We evaluate \textit{MegaAgent}'s capabilities through a series of experiments. While we first establish its performance on standard benchmarks, including reasoning and math problems (§\autoref{benchmark-exp}), our primary focus is on two scenarios demanding extensive multi-agent coordination: software development and social simulation. These latter tasks are chosen because they provide a more realistic representation of coordination challenges in human societies, often exceeding the capacity of a single LLM agent \cite{chen2023agentverse, guo2024large}.

\noindent We focus on the following three research questions:

\noindent \textbf{RQ1:} Can \textit{MegaAgent} perform well on standard benchmarks compared with other baselines? (§\autoref{benchmark-exp})

\noindent \textbf{RQ2:} Can \textit{MegaAgent} complete a task requiring extensive coordination without a predefined SOP? How do other baselines compare? (§\autoref{soft-exp})

\noindent \textbf{RQ3:} Can \textit{MegaAgent} be effectively scaled to handle more complex tasks that involve a significantly larger number of agents, showcasing its scalability? How does it compare to other baselines?(§\autoref{policy-experiment})

\subsection{RQ1: Standard Benchmarks} \label{benchmark-exp}
To evaluate the performance of \textit{MegaAgent} on standard benchmarks, we select the following tasks commonly used in previous MAS.

\subsubsection{Experiment Setup}
We conduct experiments using several standard benchmarks widely adopted in prior MAS research \cite{zhang2024agentprune, zhang2025evoflow}. These benchmarks include: (1) MBPP \cite{austin2021program}; (2) HumanEval \cite{chen2021evaluating}; (3) MATH \cite{hendrycks2021measuring} and (4) GSM-8K \cite{cobbe2021training}. For all experiments, \textit{MegaAgent} and baseline models utilize the GPT-4o API to ensure a fair comparison. Performance is assessed based on quantitative accuracy, measuring the percentage of correctly solved problems for each benchmark. 

\subsubsection{Experiment Results and Analysis}
The detailed results of our comparative evaluation are presented in Table~\ref{tab:standard_benchmarks}.

\begin{table}[htbp]
\centering
\resizebox{\linewidth}{!}{%
\begin{tabular}{ccccc}
\toprule
\rowcolor{champagne} 
\textbf{Model} & \textbf{MBPP} & \textbf{HumanEval} & \textbf{MATH} & \textbf{GSM-8k} \\
\midrule
\rowcolor{gray!10}MetaGPT & 81.7 & 82.3 & N/A & N/A \\
Camel & 78.1 & 57.9 & 22.3 & 45.6 \\
\rowcolor{gray!10}AgentVerse & 82.4 & 89.0 & 54.5 & 81.2 \\
AutoGen & 85.3 & 85.9 & \textbf{69.5} & 87.8 \\
\rowcolor{gray!10}MegaAgent & \textbf{92.2} & \textbf{93.3} & 69.0 & \textbf{93.0} \\
\bottomrule
\end{tabular}%
}
\caption{Performance comparison on standard benchmarks. All models utilized the GPT-4o API. Results are reported as accuracy (\%). N/A indicates that the benchmark is not applicable (e.g., MetaGPT primarily supports code generation tasks and was not evaluated on MATH/GSM-8K in the compared sources).}
\label{tab:standard_benchmarks}
\end{table}

As shown in Table~\ref{tab:standard_benchmarks}, \textit{MegaAgent} consistently outperforms the baseline models—MetaGPT, Camel, AgentVerse, and AutoGen—across all evaluated benchmarks. For instance, on MBPP and HumanEval, \textit{MegaAgent} achieves accuracies of 92.2\% and 93.3\% respectively, surpassing the next best performing model, AutoGen. Similarly, in the more reasoning-intensive tasks of MATH and GSM-8K, \textit{MegaAgent} demonstrates superior performance, achieving 69.0\% on MATH (comparable to AutoGen but with higher scores on other benchmarks) and a leading 93.0\% on GSM-8K. This consistent outperformance suggests that \textit{MegaAgent}'s can perform well on standard benchmarks.

\subsection{RQ2: Gobang Game Development} \label{soft-exp}
Gobang, also known as Gomoku, is a strategic board game played between two participants who take turns placing black and white pieces on a grid. The objective is to be the first to align five consecutive pieces horizontally, vertically, or diagonally\footnote{\url{https://en.wikipedia.org/wiki/Gomoku}}. We select game development as a test scenario because it effectively evaluates an MAS' coding and coordination abilities. The task requires generating both backend logic and frontend components while involving extensive collaboration among roles like product  manager, game logic designer, and software developers. This setting provides a robust evaluation of \textit{MegaAgent}’s capabilities in coordination, autonomy, and parallelism in a project.

\subsubsection{Experiment Setup}
We conduct this experiment using the GPT-4o API\footnote{\url{https://openai.com/index/hello-gpt-4o/}}, setting the 'temperature' parameter to $0$ to ensure more deterministic responses \cite{achiam2023gpt}. The experiment begins by feeding the meta prompt to \textit{MegaAgent} shown in \autoref{fig:softwareclub}. More details are in \autoref{megaagent-gobang}.

% \textit{You are Bob, the leader of a software development club. Your club's current goal is to develop a Gobang game with an AI, and can be executed by running 'main.py'.}

\begin{figure}[htp]
\centering
\begin{tikzpicture}
    \definecolor{chatcolor1}{HTML}{FFFACD}
    \definecolor{shadecolor}{gray}{0.95}

    % Define the first node
    \node[
        align=left,
        text width=0.45\textwidth,
        fill=shadecolor,
        rounded corners=1mm,
        draw=black,
        line width=0.5mm
    ] (node1) at (0,0) {
        \textit{You are Bob, the leader of a software development club. Your club's current goal is to develop a Gobang game with an AI, and can be executed by running \texttt{'main.py'}.}
    };

\end{tikzpicture}
\caption{Gobang Game Meta Prompt}
\label{fig:softwareclub}
\end{figure}

% \textcolor{red}{why they can be run in parallel, which you can add the details in the section 3.2. The section 3.2 should be refactored.}

For comparative analysis, we employ AutoGen, MetaGPT, CAMEL, and AgentVerse to perform the same task. We manually adjust their backbones to GPT-4o or GPT-4 when GPT-4o is incompatible with their configurations. To ensure a fair evaluation, we design prompts tailored to each baseline's requirements while adhering to the guidelines specified in their respective papers to determine appropriate testing methods. Further details are in \autoref{sec-gobangbaseline}.

% If a baseline's LLM-MA system includes programming examples, we use those examples as a basis, modifying only the final objective to match our Gobang task, while keeping other predefined SOPs unchanged. Otherwise, we input the prompt according to the methodology provided in their demonstration. For our own system, we add one additional prompt for role-playing, which is part of our system architecture and remains consistent regardless of the final objective. These approaches allow us to maintain the integrity of each system's design while adapting it to our specific objective of developing the Gobang game. 
% \textcolor{blue}{Qian: here, reviewers may argue we do not use the same prompt, so it is not fair. Can you add an explanation that we follow the guidance of each baseline's paper to decide the prompt?}

\subsubsection{Evaluation Metrics}
To evaluate the generated Gobang game, we establish the following evaluation metrics: \textbf{(1) Error-Free Execution}, which assesses the program's ability to run without errors; \textbf{(2) User Move}, which evaluates the user's ability to make a move; \textbf{(3) AI Move}, which measures the AI's ability to make a move; and \textbf{(4) Game Termination}, which ensures the game's ability to end correctly when there are five consecutive pieces.

% \item AI Competency: The AI's strategic strength and effectiveness.
% \item Additional Functionalities: Features such as Undo \& Redo, Save \& Load, and Quit \& Restart.
% \end{enumerate}

% \textcolor{red}{5,6 are not necessary here as our product cannot finish the 5 and 6}
% \textcolor{red}{I think a well-defined evaluation metric is necessary here. We can say like, whether the output system can imitate the gobang actions, etc...}

\subsubsection{Experiment Results}
As shown in \autoref{tab:caseresult}. \textit{MegaAgent} autonomously generates an SOP involving seven agents, effectively coordinates their tasks, and successfully develops a fully functional Gobang game with an interactive interface within 800 seconds. \textbf{These achievements fulfill all task requirements, making \textit{MegaAgent} the only system capable of producing a complete and operational game, unlike baseline models that either produce incomplete results or fail entirely.} Further details are provided in \autoref{megaagent-gobang}. The performance of other baseline models is analyzed below:

\begin{table*}[!ht]
\centering
\resizebox{\textwidth}{!}{%
\begin{tabular}{c|cccccc|c}
\toprule
\rowcolor{champagne} 
\textbf{Model} & \textbf{Error-Free Execution} & \textbf{User Move} & \textbf{AI Move} & \textbf{Game Termination} & \textbf{\# of Agents} & \textbf{Time(s)} & \textbf{Time/Agent (s)} \\
\midrule
\rowcolor{gray!10}AutoGen & \textcolor{green(pigment)}{\ding{51}} & \textcolor{green(pigment)}{\ding{51}} & \textcolor{darksalmon}{\ding{55}} & \textcolor{darksalmon}{\ding{55}} & 2 & 180 & 90 \\
MetaGPT & \textcolor{green(pigment)}{\ding{51}} & \textcolor{green(pigment)}{\ding{51}} & \textcolor{darksalmon}{\ding{55}} & \textcolor{darksalmon}{\ding{55}} & 6 & 480 & 80 \\
\rowcolor{gray!10}CAMEL & \textcolor{darksalmon}{\ding{55}} & \textcolor{darksalmon}{\ding{55}} & \textcolor{darksalmon}{\ding{55}} & \textcolor{darksalmon}{\ding{55}} & 2 & 1,830 & 915 \\
AgentVerse & \textcolor{darksalmon}{\ding{55}} & \textcolor{darksalmon}{\ding{55}} & \textcolor{darksalmon}{\ding{55}} & \textcolor{darksalmon}{\ding{55}} & 4 & 1,980 & 495 \\
\midrule
\rowcolor{gray!10}\textbf{MegaAgent} & \textcolor{green(pigment)}{\ding{51}} & \textcolor{green(pigment)}{\ding{51}} & \textcolor{green(pigment)}{\ding{51}} & \textcolor{green(pigment)}{\ding{51}} & 7 & 800 & 114 \\
\bottomrule
\end{tabular}
}
\caption{Gobang Game Development Results}
\label{tab:caseresult}
\end{table*}

\noindent \textbf{AutoGen:}  
AutoGen employs two agents but fails to produce a valid game move. After approximately three minutes, it generates a program ending with \texttt{\# To be continued..} and becomes stuck when attempting execution. The likely cause of this failure is its overly simplistic SOP, lacking critical inter-agent communication steps such as code review. More details are provided in \autoref{autogen-gobang}.

\noindent \textbf{MetaGPT:}  
Despite generating six agents, MetaGPT fails to produce a functional AI move in any trial. The main issues include: (1) unexecutable code due to the lack of debugging tools, (2) incorrect program generation, such as creating a \textit{tic-tac-toe game}\footnote{\url{https://en.wikipedia.org/wiki/Tic-tac-toe}} instead of a Gobang game, likely due to a simplistic SOP and insufficient agent communication, and (3) infinite loops caused by incomplete implementations. More details are in \autoref{metagpt-gobang}.

\noindent \textbf{CAMEL:}  
CAMEL cannot produce executable Python code using two agents, likely due to weak planning and limited contextual reasoning capabilities. More details are in \autoref{camel-gobang}.  

\noindent \textbf{AgentVerse:}  
AgentVerse generates four agents to complete the task but faces significant issues. In the first two trials, the agents repeatedly reject results for all ten rounds. In the third trial, while the result is accepted, the generated code contains numerous placeholders and remains unexecutable. The likely cause of failure is an overly rigid task outline during the planning stage, which current LLMs struggle to fulfill. More details are in \autoref{agentverse-gobang}.

\subsubsection{Ablation Study}
To validate the necessity of each component design in \textit{MegaAgent}, we conduct an ablation study, with results in \autoref{tab:gobang-ablation}.

\begin{table}[!h]
\centering
\resizebox{\linewidth}{!}{%
\begin{tabular}{c|c|c|c|c}
\toprule
\rowcolor{champagne} 
\textbf{Components} & \textbf{Completed Metrics} & \textbf{\# Agents} & \textbf{Time(s)} & \textbf{Time/Agent (s)} \\
\midrule
\rowcolor{gray!10}Full & (1) (2) (3) (4) & 7 & 800 & 114 \\
w/o hierarchy & (1) (2) & 5 & 920 & 184 \\
\rowcolor{gray!10}w/o parallelism & (1) (2) (3) (4) & 7 & 4,505 & 643 \\
w/o monitoring & (1) (2) (3) & 7 & 300 & 42\\
\bottomrule
\end{tabular}
}
\caption{Gobang Ablation Study Results}
\label{tab:gobang-ablation}
\end{table}

Removing the hierarchical structure reduces agent usage to 5 but increases completion time to 920 seconds while achieving only basic metrics. Without parallelism, task groups complete their tasks sequentially, increasing time complexity from $O(\log n)$ to $O(n)$, which raises the execution time per agent from 114 seconds to 643 seconds. Removing monitoring reduces execution time to 300 seconds but fails to meet essential metrics. These findings underscore that parallel execution, hierarchy, and monitoring are all crucial for both task completion and execution speed. More details are in \autoref{gobang-ablation}.

% \subsubsection{Ablation Study}

% \textcolor{red}{add parallel vs non-parallel time}
% \textcolor{blue}{Qian: you can add code screenshots and the game screenshots here. This is the most important we want to show to reviewers.}

% \subsubsection{Ablation Test}
% To demonstrate the effectiveness of our task-splitting technique, we manually designed an SOP for Gobang game development while keeping other modules unchanged. This resulted in a program that closely resembles the original MegaAgent's output. It shows that MegaAgent-generated SOPs have comparable performance to predefined SOPs. The full hand-written SOP for this task can be found in our supplementary material.

% \textcolor{blue}{Qian: As we have more space, you can add the pre-defined SOPs' examples here}

% \textcolor{red}{compare SOP}

\subsubsection{Cost Analysis}
To evaluate token usage and better understand the efficiency of the Gobang game generation, we provide a detailed cost analysis. The analysis is divided into three stages: Planning, Task-Solving, and Merging, each representing distinct phases of the system's operation. The Planning stage focuses on initial strategy generation, the Task-Solving stage handles the core game-solving computations, and the Merging stage consolidates results for final outputs. We have two key insights from the results in \autoref{tab:cost_analysis} as follows:

\begin{table}[ht]
\centering
\resizebox{0.49\textwidth}{!}{%
\begin{tabular}{c|cccc}
\toprule
\rowcolor{teal!25} 
\textbf{Stage} & \textbf{\# Input Tokens} & \textbf{\# Output Tokens} & \textbf{\# Total Tokens} & \textbf{Time (s)} \\ 
\midrule
\rowcolor{gray!10}Planning & 42,947 & 12,347 & 55,294 & 0--60 \\ 
Task-Solving & 1,098,573 & 55,022 & 1,153,595 & 30--840 \\ 
\rowcolor{gray!10}Merging & 22,099 & 1,493 & 23,592 & 840--870 \\ 
\midrule
Total & 1,163,619 & 68,862 & 1,232,481 & 870 \\ 
\bottomrule
\end{tabular}
}
\caption{Token usage analysis across different stages of Gobang GPT-4o experiments.}
\label{tab:cost_analysis}
\end{table}

% \begin{table}[ht]
% \centering
% \resizebox{0.5\textwidth}{!}{%
% \begin{tabular}{c|cccc}
% \toprule
% \rowcolor{champagne} \textbf{Stage}        & \textbf{Time (s)} & \textbf{\# Input Tokens} & \textbf{\# Output Tokens} & \textbf{\# Total Tokens} \\ 
% \midrule
% \rowcolor{gray!10}Planning              & 0--60             & 42,947                & 12,347                 & 55,294                \\ 
% Task-Solving          & 30--840           & 1,098,573             & 55,022                 & 1,153,595             \\ 
% \rowcolor{gray!10} Merging               & 840--870          & 22,099                & 1,493                  & 23,592                \\ 
% \midrule
% Total        & 870                & 1,163,619    & 68,862        & 1,232,481   \\ 
% \bottomrule
% \end{tabular}
% }
% \caption{Token usage analysis across different stages of Gobang GPT-4o experiments. }
% \label{tab:cost_analysis}
% \end{table}

\noindent \textbf{Insight 1: High Resource Consumption in the Task-Solving Stage.} \label{insight1} The majority of the time and token usage occurs during the task-solving stage. This indicates that the task is inherently complex, requiring significant coordination among agents to generate solutions. This highlights the computational intensity of multi-agent interactions in solving strategic problems.

\noindent \textbf{Insight 2: Disproportionate Input and Output Token Usage.} The input token count is substantially higher than the output token count, revealing significant room for optimization in token usage. Notably, the input tokens predominantly originate from dialogues between agents. This suggests that improving the efficiency and structure of inter-agent communication could be a valuable research direction to enhance overall efficiency.

% \subsection{Planning: Travel Planner}
% We propose a popular planning experiment evaluating the plannning abilities 

\subsection{RQ3: National Policy Generation} \label{policy-experiment}
We propose a more challenging experiment: formulating national policies, which requires numerous agents to perform various tasks in complex domains such as education, health, and finance. We select this experiment because social simulations with MAS require numerous agents—potentially scaling to hundreds—to mimic a human-like society. This experiment can evaluate \textit{MegaAgent}'s autonomy, scalability, and coordination capabilities.

\subsubsection{Experiment Setup}
Due to budget constraints, we use the GPT-4o-mini API for this experiment conducted by \textit{MegaAgent}. For comparative analysis, we utilize AutoGen, MetaGPT, CAMEL, and AgentVerse to perform the same task. We manually adapt their backbone LLMs to GPT-4o or GPT-4 when GPT-4o is incompatible with their code configurations. The meta prompt we feed into \textit{MegaAgent} is shown in \autoref{fig:nationLeader}, with more details provided in \autoref{megaagent-national}. Descriptions of the other baseline settings are included in \autoref{national-policy-baseline}.

\begin{figure}[htp]
\centering
\begin{tikzpicture}
    \footnotesize
    \definecolor{chatcolor1}{HTML}{FFFACD} % Highlight color for the prompt
    \definecolor{shadecolor}{gray}{0.95} % Background color for context
    \fontfamily{cmss}\selectfont
    % Formatting for the user's prompt
    \node[align=left, text width=0.45\textwidth, fill=shadecolor, rounded corners=1mm, draw=black, line width=0.5mm, fill opacity=0.7, text opacity=1, anchor=north west] (node1) at (0,0) {
        \textit{You are NationLeader, the leader of a pioneering nation. You want to develop the best detailed policy for your cutting-edge country in `policy\_department.txt'. You are now recruiting ministers and assigning work to them. For each possible minister, please write a prompt.}
    };
    % Placeholder for generated responses or annotations    
\end{tikzpicture}
\vspace{-5pt}
\caption{National Policy Generation Meta Prompt}
\label{fig:nationLeader}
\end{figure}

% The full prompt is in the appendix.

% \textcolor{blue}{Qian: You can add some prompts here. Then, you say But due to the space limit, the further prompts are in our code repository. }

\subsubsection{Evaluation Metrics}
To assess \textit{MegaAgent}'s national policy reliability, we implement the \textit{LLM-as-a-Judge} framework using five advanced models: Claude-3.5\footnote{\url{https://www.anthropic.com/claude/sonnet}}, GPT-4o-mini, GPT-4o, o1-mini, and o1-preview \cite{achiam2023gpt}. Our validation dataset contains 50 authentic national policies and 50 non-policy texts (multi-turn conversations, meeting summaries) \cite{zheng2023judging} to test policy discrimination capability. Using the standardized prompt in \autoref{fig:policyEvaluation}, models achieve 89\% average identification accuracy in \autoref{tab:evaluation_results}, confirming evaluation validity. Complete validation details are in \autoref{national-valid}.

% \subsubsection{Evaluation Metrics}
% To evaluate the reliability of \textit{MegaAgent}'s generated national policies, we use the \textit{LLM-as-a-Judge} framework to assess their reasonableness. We select five advanced and widely recognized LLMs: Claude-3.5\footnote{\url{https://www.anthropic.com/claude/sonnet}}, gpt-4o-mini, gpt-4o, o1-mini, and o1-preview \cite{achiam2023gpt} for this evaluation.

% To validate the LLMs' ability to assess national policies, we create a validation dataset containing both authentic national policies and various unrelated text formats \cite{zheng2023judging}. This setup tests whether the LLMs can distinguish real policies from non-policy texts. We use the evaluation prompt presented in \autoref{fig:policyEvaluation} for all selected LLMs. The results in \autoref{tab:evaluation_results} show that, on average, the models achieve an 89\% accuracy rate in identifying real national policies, demonstrating their effectiveness in this evaluation framework. Additional details are provided in \autoref{national-valid}.

% For the actual evaluation, we apply the same prompt from \autoref{fig:policyEvaluation} across all models. The prompt asks the models to assess the reasonableness of the policies generated by \textit{MegaAgent} as national policies.

\begin{figure}[htp]
\centering
\begin{tikzpicture}
    \definecolor{chatcolor1}{HTML}{FFFACD}
    \definecolor{shadecolor}{gray}{0.95}

    % Define the first node
    \node[
        align=left,
        text width=0.45\textwidth,
        fill=shadecolor,
        rounded corners=1mm,
        draw=black,
        line width=0.5mm
    ] (node1) at (0,0) {
        \textit{"Is this policy reasonable as a national policy? Please return your answer with clear nuances: Agree, Disagree, or Neutral with detailed explanations."}
    };

\end{tikzpicture}
\caption{National Policy Evaluation Prompt}
\label{fig:policyEvaluation}
\end{figure}

\subsubsection{Experiment Results}
\begin{figure}[htbp]
    \centering
    \includegraphics[width=\linewidth]{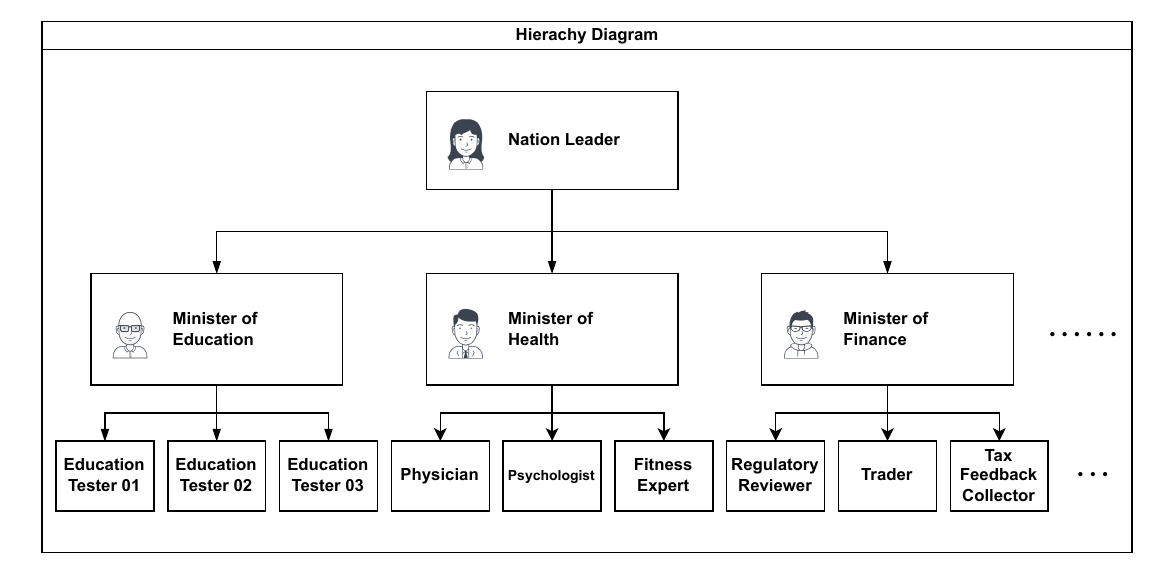}
    \caption{MegaAgent's Generated National Policy Structure}
    \label{fig:hierarchy}
\end{figure}

We present National Policy Generation's experimental results in \autoref{tab:industry}. It shows \textit{MegaAgent}'s ability to generate complete and reasonable policies using a significantly larger number of agents within competitive time limits. The results show that \textbf{\textit{MegaAgent} outperforms baseline models by producing complete policies with 590 agents in 2,991 seconds}. Notably, \textit{MegaAgent}'s average processing time per agent is 5 seconds, significantly faster than the best-performing baseline at 40 seconds per agent, demonstrating its scalability. The structure of the policies generated by \textit{MegaAgent} is illustrated in \autoref{fig:hierarchy}, with detailed outputs provided in \autoref{megaagent-result}.

\begin{table}[ht]
\centering
\resizebox{0.49\textwidth}{!}{%
\begin{tabular}{c|ccc|c}
\toprule
\rowcolor{champagne}\textbf{Model} & \textbf{Outputs} & \textbf{\# Agents} & \textbf{Time (s)} & \textbf{Time/Agent (s)} \\
\midrule
\rowcolor{gray!10}AutoGen  & Outline          & 1                      & 40               & 40 \\
MetaGPT  & Python Program   & 6                      & 580              & 97 \\
\rowcolor{gray!10}CAMEL       & Plans            & 2                      & 1,380            & 690 \\
AgentVerse  & None         & 4                      & 510              & 128 \\
\midrule
\rowcolor{gray!10}MegaAgent                      & Complete Policies & 590                    & 2,991            & 5 \\
\bottomrule
\end{tabular}
}
\caption{National Policy Generation Results}
\label{tab:industry}
\end{table}

To evaluate the reliability of \textit{MegaAgent}'s generated national policies, we feed the prompt in \autoref{fig:policyEvaluation} to chosen advanced LLMs for reasonability assessment. As shown in \autoref{tab:real_evaluation_results}, an average of 27.4 out of 31 policies are judged as reasonable by LLMs. This result highlights \textit{MegaAgent}'s effectiveness in generating well-justified policies in this social simulation experiment.

\begin{table}[ht]
\centering
\resizebox{0.3\textwidth}{!}{%
\begin{tabular}{l|ccc}
\toprule
\rowcolor{blue!15}\textbf{Model}          & \textbf{\# Agree} & \textbf{\# Disagree} & \textbf{\# Neutral} \\ 
\midrule
\rowcolor{gray!10}Claude-3.5              & 26                  & 1                     & 4                     \\ 
gpt-4o-mini             & 28                  & 0                     & 3                     \\ 
\rowcolor{gray!10}gpt-4o              & 25                  & 2                      & 4                     \\ 
o1-mini                 & 29                  & 2                     & 0                    \\ 
\rowcolor{gray!10}o1-preview                 & 29                  & 1                     & 1                     \\ 
\midrule
\textbf{Average} & \textbf{27.4}       & \textbf{1.2}          & \textbf{2.4}            \\ 
\bottomrule
\end{tabular}
}
\caption{Evaluating the Rationality of 31 Policies Generated by MegaAgent}
\label{tab:real_evaluation_results}
\end{table}
% \textcolor{red}{Qian: Add megaagent running time.}

% \textcolor{blue}{Qian: here add a hierarchy figure: which agents are in hierarchy 1, which agents are in hierarchy 2, which agents are in hierarchy 3.}

% Finally, MegaAgent presents a comprehensive policy addressing all seventeen ministries of the simulated nation, including detailed instructions and amendments, as follows:
% \dirtree{%
%   .1 /project.
%     .2 {policy\_health.txt}.
%     .3 feedback\_health.txt.
%     .2 {policy\_education.txt}.
%   .3 feedback\_education.txt.
%   .2 ....
% }

% For example, policy\_health.txt looks like:

% \begin{lstlisting}
% # Health Policy
% ## Introduction
% This health policy aims to ensure universal healthcare access, promote preventive care, ...
% ## Objectives
% 1. Ensure universal access to healthcare services. ...
% ## Actions
% ### 1. Universal Healthcare Access ...
% \end{lstlisting}

% \textcolor{blue}{Qian: here also add some concrete txt contents, now it is only the txt names}

% The complete version is in the appendix. The autonomy and scalability of MegaAgent are demonstrated by its ability to enable agents to communicate with different objectives across multiple rounds and recruit new agents as needed. 

\subsubsection{Ablation Study}
To validate the necessity of each component design in \textit{MegaAgent}, we conduct an ablation study with results shown in \autoref{tab:industry-ablation}.

\begin{table}[ht]
\centering
\resizebox{\linewidth}{!}{%
\begin{tabular}{c|ccc|c}
\toprule
\rowcolor{champagne}\textbf{Components} & \textbf{Outputs} & \textbf{\# of Agents} & \textbf{Time (s)} & \textbf{Time/Agent (s)} \\
\midrule
\rowcolor{gray!10}Full & Complete Policies & 590 & 2,991 & 5 \\
w/o hierarchy & Incomplete Policies & 19 & 450 & 24 \\
\rowcolor{gray!10}w/o parallelism  & Incomplete Policies & \textgreater 100 &  \textgreater 14,400 & N.A. \\
w/o monitoring  & Policies with Placeholders & 50 &  667 & 13 \\
\bottomrule
\end{tabular}
}
*We terminate the execution without parallelism \\ after 14400 seconds.
\caption{National Policy Generation Ablation Study Results} 
\label{tab:industry-ablation}
\end{table}

Without hierarchy, only incomplete policies are produced within 450 seconds using 19 agents, indicating the importance of hierarchical design. Disabling parallelism entirely results in incomplete policies even after 14400 seconds, with over 100 agents continuously recruited but unable to complete tasks due to serialized processing bottlenecks. Removing monitoring generates policies with placeholders in 667 seconds using 50 agents, highlighting the need for continuous supervision for task completeness. 
Detailed outputs of these ablation studies are in \autoref{national-policy-ablation}.

These findings underscore that parallelism is not \textbf{merely beneficial but critical} for managing complex tasks in MAS.

\subsubsection{Cost Analysis}
To assess the token and time costs of this experiment, we perform a detailed analysis of token usage and execution time across three stages: Planning, Task-Solving, and Merging. The results are presented in \autoref{tab:policy_analysis}.

\begin{table}[ht]
\centering
\resizebox{0.49\textwidth}{!}{%
\begin{tabular}{c|cccc}
\toprule
\rowcolor{teal!25}\textbf{Stage} & \textbf{Input Tokens} & \textbf{Output Tokens} & \textbf{Total Tokens} & \textbf{Time (s)} \\ 
\midrule
\rowcolor{gray!10}Planning & 111,601 & 24,103 & 135,704 & 0--180 \\ 
Task-Solving & 8,003,124 & 343,670 & 8,346,794 & 20--2,950 \\ 
\rowcolor{gray!10}Merging & 348,264 & 13,280 & 361,544 & 2,400--3,000 \\ 
\midrule
Total & 8,463,989 & 381,053 & 8,845,042 & 3,000 \\ 
\bottomrule
\end{tabular}
}
\caption{Token usage analysis for National Policy Generation.}
\label{tab:policy_analysis}
\end{table}

% \subsection{Travel Planner Experiment}
% \textcolor{blue}{In this experiment, we employ the travelplanner dataset to test the performance}

% \begin{table}[ht]
% \centering
% \resizebox{0.49\textwidth}{!}{%
% \begin{tabular}{c|cccc}
% \toprule
% \rowcolor{champagne}\textbf{Stage}        & \textbf{Time (s)} & \textbf{Input Tokens} & \textbf{Output Tokens} & \textbf{Total Tokens} \\ 
% \midrule
% \rowcolor{gray!10}Planning              & 0--180            & 111,601               & 24,103                 & 135,704               \\ 
% Task-Solving          & 20--2,950         & 8,003,124             & 343,670                & 8,346,794             \\ 
% \rowcolor{gray!10}Merging               & 2,400--3,000      & 348,264               & 13,280                 & 361,544               \\ 
% \midrule
% Total        & 3,000                & 8,463,989    & 381,053      & 8,845,042   \\ 
% \bottomrule
% \end{tabular}
% }
% \caption{Token usage analysis for National Policy Generation. Results may vary due to network conditions and the inherent randomness of LLMs.}
% \label{tab:policy_analysis}
% \end{table}

Similar to analysis in \ref{insight1}, we observe from \autoref{tab:policy_analysis} that significant resource consumption during the task-solving stage, which dominates both time and token usage. A comparison of input-to-output token ratios between the experiments reveals consistent inefficiencies, with the first experiment showing a ratio of approximately 23:1, while the current experiment is slightly higher at 25:1. This increase suggests that the \textbf{policy generation task required additional resources for inter-agent dialogues and greater context management, likely due to the involvement of a larger number of agents.} These findings highlight the critical need to optimize token usage and enhance dialogue efficiency, which could significantly reduce resource consumption and improve overall performance in MAS.

\subsection{Scalability Analysis}
In \textit{MegaAgent}, for \(n\) agents, the hierarchical layer-to-layer communication cost is \(O(\log n)\), as agent groups at the same level operate in parallel, as illustrated in \autoref{fig:hierarchy}. In contrast, existing frameworks exhibit linear running time growth \(O(n)\) as they run serially, which becomes impractical with the number of LLM agents increasing much. The analysis is supported by our national policy generation experiment in \autoref{policy-experiment}, where \textit{MegaAgent}'s average processing time per agent is 5 seconds, compared to CAMEL's average of 700 seconds per agent. These results highlight \textit{MegaAgent}'s scalability and practicality for autonomous coordination in large-scale MAS.

%% file: sec-relatedwork.tex
\section{Related Work}
\label{sec:related-work}
We discuss the most related work here and leave more details in \autoref{additional-related-work}.

\subsection{LLM-based Multi-Agent Systems}
With the emergence of powerful LLMs \cite{achiam2023gpt, team2023gemini}, recent research on LLM-based multi-agent systems (MAS) has investigated how multiple agents can accomplish tasks through coordination, utilizing elements such as personas \cite{chen2024persona, chan2024scaling}, planning \cite{chen2023put, zhang2024timearena, yuan2023distilling}, and memory \cite{zhang2023memory, hatalis2023memory}. Unlike systems relying on a single LLM-based agent, multi-agent systems demonstrate superiority in tackling challenging tasks. Recent works, such as MetaGPT \cite{hong2023metagpt}, AutoGen \cite{wu2023autogen}, and AgentVerse \cite{chen2023agentverse}, design specific roles to achieve a task. 

However, most popular MAS heavily rely on handcrafted prompts and expert design. For instance, MetaGPT \cite{hong2023metagpt} requires users to pre-design roles like product manager and software engineer. Another limitation is these systems utilize a sequential pipeline without considering parallel execution of agents \cite{li2023camel}. Although AgentScope \cite{pan2024very} does consider this, its implementation follows a fixed trajectory in different rounds of interaction, prohibiting changes in communication partners, thus limiting performance improvement as the number of agents scales up.  

In contrast, in the real world, when many software developers are employed, they may first work on different files simultaneously, and then focus on one specific file when difficulties are encountered, sparking creative ideas to overcome challenges by coordination.  Additionally, existing MAS are restricted by their small scale and have not been applied in large-scale scenarios with complex coordination involved. We compare current popular MAS with \textit{MegaAgent} in \autoref{tab:relate-work}. We can see from the table that \textit{MegaAgent} stands out for its high autonomy, multi-file support, parallelism, and scalability.

\subsection{SOPs in MAS}
Allocating SOPs is a common approach in designing agent profiles and tasks within MAS \cite{hong2023metagpt, huang2023agentcoder, park2023generative, zhuge2024language, shi2024learning}. These systems define SOPs for both individual agents and their communication protocols. While this method has proven effective in previous works, it has two major limitations: (1) Agents may possess unforeseen capabilities that cannot be anticipated during the human design stage but become relevant during task execution \cite{rivera2024conceptagent, sypherd2024practical, piatti2024cooperate}; (2) As the scale of MAS grows—potentially involving thousands or even billions of agents—designing SOPs manually for each agent becomes infeasible \cite{mou2024individual, pan2024very}. To address this, the design mechanism must evolve, leveraging LLMs themselves, as in the \textit{LLM-as-the-Judge} concept \cite{huang2024empirical, chen2024mllm}, allowing LLMs to autonomously generate SOPs for large-scale MAS.

%% file: sec-conclusion.tex
\section{Conclusion}
\label{sec:conclusion}
We present \textit{MegaAgent}, a large-scale autonomous LLM-based multi-agent system enabling dynamic and autonomous coordination, where users only need to provide a meta prompt at the start of the process. \textit{MegaAgent} demonstrates strong foundational capabilities by outperforming existing baselines on standard benchmarks. Furthermore, through a Gobang game software development experiment, we demonstrate \textit{MegaAgent}'s superior autonomy and coordination compared to baseline models. Our social simulation on national policy generation highlights \textit{MegaAgent}'s scalability to hundreds of agents while ensuring effective coordination. With its hierarchical and adaptive design, \textit{MegaAgent} has the potential to serve as the foundational OS for future MAS. We encourage the research community to further explore enhancing agent coordination to address the increasing demands of large-scale MAS.

\section*{Limitations}
\noindent \textbf{Planning and Communication Overhead.}  
The primary bottleneck lies in the planning and communication processes among LLM agents, particularly in translating code into prompts, managing task checklists, maintaining the framework, and debugging. As the number of agents and communication rounds increases, input-output token consumption grows substantially, affecting both efficiency and cost. Future work should explore advanced token summarization, semantic compression, and efficient dialogue storage methods.  

\noindent \textbf{Hallucination in Agent Outputs.}  
Despite using task-specific checklists to monitor agent actions, occasional hallucinations persist, with output formats sometimes deviating from expected requirements. Since the checklists themselves are generated by LLMs, errors may propagate. Addressing this requires more robust verification mechanisms, potentially involving external expert knowledge bases before, during, or after agent response generation.  

\noindent \textbf{API Cost and Model Integration.}  
\textit{MegaAgent}'s reliance on GPT-4 incurs high API costs. While cheaper alternatives exist, they may lack generalizability. A promising direction would involve integrating specialized LLMs for specific tasks, leveraging models that excel in certain domains while maintaining efficient communication and data sharing across the LLMs.

\section*{Acknowledgements}
This research is supported by the National Research Foundation, Singapore and Infocomm Media Development Authority under its Trust Tech Funding Initiative. Any opinions, findings and conclusions or recommendations expressed in this material are those of the author(s) and do not reflect the views of National Research Foundation, Singapore and Infocomm Media Development Authority. The work of Mr. Tianyu Wang and Mr. Jingsheng Liang was done when they were visiting students at National University of Singapore.

% \section*{Broader Impacts}
% \noindent \textbf{Reducing Inefficiencies in Complex Tasks.}  
% \textit{MegaAgent}'s hierarchical multi-agent structure could improve efficiency in other domains requiring complex planning and collaboration, such as legal drafting, project management, and research coordination. 

% \noindent \textbf{MAS System Design.} \textit{MegaAgent} framework redefines agent system design by treating it as an OS for large-scale MAS coordination. Its hierarchical structure mirrors modern OS principles, where admin agents supervise task execution while the \textit{Boss Agent} oversees system-wide operations. This design introduces a flexible and adaptive blueprint for future agent systems, integrating real-time monitoring, dynamic task assignment, and autonomous failure recovery.

% \noindent \textbf{Ethical and Social Considerations.}  
% The deployment of MAS in social simulations could reshape societal structures by reducing human involvement in decision-making processes. Ensuring fairness, equity, and accountability will be essential as these systems are scaled up.  Monitoring and mitigating potential misuse or bias in generated content from LLMs should be prioritized through ethical guidelines and technical safeguards.

%% file: sec-appendix.tex
\clearpage
\onecolumn
\appendix 
\etocdepthtag.toc{mtappendix}
\etocsettagdepth{mtchapter}{none}
\etocsettagdepth{mtappendix}{subsection}
\renewcommand{\contentsname}{Appendix}
\tableofcontents 
\clearpage
\twocolumn

% \section*{Appendix}
\section{Algorithm}
\begin{algorithm}[H]
\caption{Agent–Level Message‐Queue Routine \\(executed by every agent $a$)}
\label{alg:msgqueue}
\begin{algorithmic}[1]
\State \textbf{Message queues:} $Q_a\gets\emptyset$  \Comment{FIFO buffer owned by $a$}
\While{system is running}
    \If{$Q_a=\emptyset$}                                   \Comment{\textbf{Idle State}}
        \State \Call{Sleep}{$\Delta t_{\text{poll}}$}       \Comment{no token cost}
        \State \textbf{continue}
    \EndIf
    \State $B\gets$ \Call{DequeueBatch}{$Q_a$}              \Comment{\textbf{Processing State}}
    \State $R\gets$ \Call{LLM\_Infer}{$a$, $B$}
    \While{\Call{HasFuncCall}{$R$}}               %\Comment{handle tool calls}
        \State $(f,Args...)\gets$\Call{ParseCall}{$R$}
        \State $o\gets$\Call{Exec}{$f,Args...$}
        \State $B\gets$\Call{Concat}{$B,o$}
        \State $R\gets$\Call{LLM\_Infer}{$a,B$}
    \EndWhile
    \State $R'\gets$ \Call{Validate}{$R$}                   \Comment{format \& safety checks}
    \ForAll{$(m, a') \in R'$} \Comment{\textbf{Response State}}
    \State \Call{Enqueue}{$Q_{a'}$, $m$}
    \EndFor

    \State \Call{UpdateMemory}{$a,B,R'$}            \Comment{episodic/store}
    \State \Call{UpsertVecDB}{$a,B,R'$}             \Comment{vector DB}
   
    \If{$Q_a\neq\emptyset$}
        \State \textbf{continue} \Comment{next Processing State}
    \Else
        \State \textbf{goto Idle State}
    \EndIf
\EndWhile
\end{algorithmic}
\end{algorithm}

\begin{algorithm}[H]
\caption{MegaAgent Framework}
\label{alg:megaagent_revised}
\begin{algorithmic}[1]
\Require Prompt $\mathcal{P}$
\Ensure Final Deliverable $\mathcal{O}$
\State \textbf{Setup:} Instantiate root \textit{BossAgent} $A_0$
\State $A_0$ decomposes $\mathcal{P}$ into subtasks $S=\{s_1,\dots,s_n\}$
\ForAll{$s_i\in S$}
    \State $A_0$ recruits an \textit{AdminAgent} $A_i$ and enqueues task spec to $Q_{A_i}$
\EndFor
\Comment{All agents run Algorithm~\ref{alg:msgqueue} in parallel}
\Repeat
    \ForAll{\textit{AdminAgent} $A_i$ \textbf{in parallel}}
        \State $A_i$ splits $s_i$ if needed and enqueues sub-tasks to new or existing worker queues
        \State $A_i$ monitors outputs returned through its queue; on failure, retries or hires extra agents
    \EndFor
    \State \textbf{Synchronise:} Admin agents exchange dependency results via their queues
\Until{\textbf{all} queues are empty \textbf{and} every task validated}
\State $A_0$ aggregates verified outputs $\rightarrow \mathcal{O}$
\State \Return $\mathcal{O}$
\end{algorithmic}
\end{algorithm}

\section{Experimental Environment}
All experiments are conducted using an NVIDIA A100-80G Tensor Core GPU, utilizing Tier 5 APIs for both ChatGPT-4o and ChatGPT-4o mini \footnote{\url{https://platform.openai.com/docs/guides/rate-limits/usage-tiers?context=tier-five}}.

\section{Supplementary Related Work} \label{additional-related-work}

\subsection{Multi-Agent Coordination}
Recent advances in LLMs \citep{achiam2023gpt, team2023gemini} have spurred the creation of multi-agent frameworks where specialized models collaborate on intricate tasks like code synthesis, mathematical analysis, and decision-making \citep{hong2023metagpt, chen2023agentverse}. Innovations in role specialization \citep{chen2024persona, chan2024scaling}, adaptive planning \citep{yuan2023distilling, luo2024multi}, and knowledge retention \citep{zhang2023memory, hatalis2023memory, luo2024ai} enable these systems to surpass single-agent performance. Platforms such as MetaGPT \citep{hong2023metagpt} and AgentVerse \citep{chen2023agentverse} exemplify effective coordination strategies through structured role assignment.

\subsection{Reasoning and Planning}
\textbf{Reasoning.} The Chain-of-Thought (CoT) reasoning framework illustrates that detailed reasoning conducted in multiple stages is notably beneficial for the effectiveness of LLMs, particularly when compared to the limitations of single-step reasoning~\citep{CoT}. Multi-stage reasoning significantly enhances LLM performance by decomposing problems into sequential steps \citep{CoT}, mirroring human cognition. Single-step approaches often neglect critical intermediate phases, reducing solution accuracy \citep{CoT}.

While CoT \citep{CoT} employs one LLM invocation, newer techniques integrate search algorithms like MCTS \citep{Decoding}, Q-star search~\citep{chakraborty2024transfer} and recursive validation \citep{fubreak} during token generation to refine outputs. Parallel LLM invocations \citep{brown2024large} and answer selection via CoT-SC \citep{wang2022self} improve reliability. Structured approaches like ToT \citep{ToT} and GoT \citep{GoT} organize reasoning paths graphically, while collaborative agent simulations enable decentralized problem-solving \citep{li2023camel, hong2024metagpt}. And some research indicates that integrating knowledge graphs~\citep{luoreasoning, sunthink, zhang2023live} and prompt structuring~\citep{jiang2023structgpt} can empower LLMs reasoning ability. Multiple agents is another form of using parallel LLM to  enhance reasoning performance and solve problems~\citep{hu2023bert4eth,li2023camel, hong2024metagpt, liang2023encouraging, duimproving}.

\textbf{Planning.} Complex task resolution and reasoning requires divide-and-conquer strategies, where LLMs act as coordinators partitioning problems and delegating subtasks \citep{hong2024metagpt, wangvoyager}. Hybrid systems combining neural and symbolic reasoning further enhance planning efficiency \citep{RecursiveReasoning}. Some recent research begins utilizing LLMs to independently orchestrate planning and scheduling~\citep{hong2024metagpt, wu2024autogen, zhoulanguage, wangvoyager,RecursiveReasoning}. 

\subsection{External Memory and Tools}

\textbf{External Tools.}
Modern LLMs excel at invoking external APIs for tasks requiring web search \citep{ToolLLM}, computation \citep{NEURIPS2023_d842425e}, or system operations \citep{AIOS}. In the practical applications, the external function descriptions are sent to LLMs to allow them to choose suitable functions to execute. \citep{Granite-Function}. For example, numerical functions APIs can be particularly effective for arithmetic tasks \citep{pmlr-v202-gao23f, NEURIPS2023_e3936777}. And with the Internet search, LLMs can receive more information to enhance its task solution performance~\citep{yao2022react, FreshLLMs}.

\textbf{External Memory.} 
Incorporating external data sources like research papers or databases \citep{PopQA} reduces factual inaccuracies. Domain-specific enhancements in legal \citep{pipitone2024legalbench}, medical \citep{jeong2024improving}, and financial \citep{li-etal-2024-alphafin} contexts demonstrate significant performance gains.

\subsection{Operational Protocols in Multi-Agent Systems}
Allocating SOPs is a common approach in designing agent profiles and tasks within MAS \citep{hong2023metagpt, huang2023agentcoder, park2023generative, zhuge2024language, shi2024learning}. These systems define SOPs for both individual agents and their communication protocols. While this method has proven effective in previous works, it has two major limitations: (1) Agents may possess unforeseen capabilities that cannot be anticipated during the human design stage but become relevant during task execution \citep{rivera2024conceptagent, sypherd2024practical, piatti2024cooperate}; (2) As the scale of MAS grows—potentially involving thousands or even billions of agents—designing SOPs manually for each agent becomes infeasible \citep{mou2024individual, pan2024very}.

\subsection{LLM Inference and Serving}
\textbf{Prefilling and Decoding.} Modern language model deployment relies on two distinct operational phases. During context initialization, the system processes input prompts to establish attention key-value states – a memory-saving process that prevents redundant recomputation of historical token embeddings \citep{pope2023efficiently}. This phase is particularly critical for applications requiring sequential coherence, such as interactive chatbots.

The subsequent token generation phase produces output sequences using strategies like deterministic beam search \citep{pryzant2023automatic} or stochastic sampling \citep{brown2024large}. Beam search balances quality and computational cost, while temperature-controlled sampling introduces controlled randomness for creative outputs.

\textbf{Memory Optimization Techniques.}
Efficient KV cache management has become pivotal for high-throughput LLM serving. The PagedAttention methodology \citep{kwon2023efficient} revolutionized this domain by organizing cached attention states into non-contiguous memory blocks, enabling dynamic allocation proportional to sequence lengths. Leading inference engines like vLLM \citep{kwon2023efficient}, TensorRT-LLM \citep{tensorrtllm}, and HuggingFace TGI \citep{tgi} have implemented this approach with varying architectural adaptations.

LLM serving has seen a surge of research activity in recent years, with many systems developed to address the different challenges. The systems include TensorFlow Serving~\citep{olston2017tensorflow}, Clockwork~\citep{gujarati2020clockwork}, AlpaServe~\citep{li2023alpaserve},  Orca~\citep{orca}, vllm~\citep{kwon2023vllm}, SGLang~\citep{zheng2023sglang} and others. These serving system explore many aspects including batching, caching, placement, scheduling, model parallelism for the serving of single or multiple models. 

\textbf{Serving Metrics.}
System designers evaluate LLM serving efficiency through multiple quantitative measures: (1) TTFT (Time To First Token) that is critical for user-perceived responsiveness; (2) TPOT (Time Per Output Token)that impacts streaming experience quality; (3) Throughput that measures total tokens processed per unit time; (4) SLO Compliance Rate which is the percentage of requests meeting latency guarantees. Different serving frameworks utilize different serving strategies to optimize them.

\textbf{Scheduling Requests.}  
Effective scheduling requires balancing conflicting objectives: maximizing GPU utilization through batching while maintaining strict latency constraints \citep{ali2020batch}. Advanced systems employ: (1) Dynamic Batching: Groups requests by similar context lengths; (2) Iterative Scheduling: Interleaves processing of new and ongoing generations; (3) Predictive Scaling: Anticipates resource needs via request length estimation \citep{zheng2024response}. Modern schedulers like FastServe \citep{wu2023fast} implement priority queues based on input complexity, while response-aware systems \citep{jin2023s} optimize for tail latency reduction. The industry-wide adoption of continuous batching allows incremental addition of requests to active computation batches, dramatically improving hardware utilization compared to static batching approaches.

\subsection{System Optimization of MAS}

\textbf{Workflow Composition.}
Agent-based applications typically involve interconnected LLM calls organized as computational graphs. These directed acyclic graphs (DAGs) represent data dependencies between model invocations, where nodes correspond to individual agents and edges define execution prerequisites \citep{langchain}. Tools like Microsoft's PromptFlow \citep{promptflow} provide visual editors for constructing such workflows, enabling latency prediction and parallel execution planning.

\textbf{Legacy Graph Processing.}
Previous generation DAG schedulers \citep{tez,dryad,sonic} focused on general data processing optimizations: (1) Pipeline parallelism for multi-stage workloads; (2) Data locality-aware task placement; (3) Inter-node communication minimization. While effective for traditional dataflows, these systems lack semantic understanding of LLM-specific patterns like prompt reuse opportunities or attention cache sharing potential across requests in MAS.

\textbf{DAG-aware LLM Serving.} Recent innovations address LLM workflow peculiarities through two complementary approaches.

\textit{Semantic Task Chaining.} Parrot \citep{lin2024parrot} introduces reusable context containers (semantic variables) that explicitly track dependencies between LLM calls. This enables cross-request prompt deduplication and context-aware scheduling, reducing redundant computation by up to 40\% in benchmark tests.

\textit{Cross-Request Cache Optimization.} SGLang \citep{zheng2023sglang} implements RadixAttention – a prefix-aware KV cache management system. By maintaining an LRU-cached radix tree of attention states, the framework identifies and reuses common prompt prefixes across concurrent requests. When combined with cache-aware scheduling, this technique demonstrates 3.8× throughput improvement on workflows with shared context segments.

These advancements highlight the necessity of specialized scheduling mechanisms for LLM agent systems, contrasting with the one-size-fits-all approaches of conventional DAG processors.

\subsection{LLM-based Agents Coordination}
The coordination between LLM-based agents is critical infrastructure for supporting MAS \cite{guo2024large}. There are three main coordination paradigms: cooperative, debate, and competitive. \textit{MegaAgent} focuses on the coordination paradigm, aiming to have agents work together toward a shared goal. Within the cooperative paradigm are three main structures: layered, decentralized, and centralized. Layered communication is organized hierarchically, with agents at each level having distinct roles and each layer interacting with adjacent layers \cite{liu2023dynamic}. Decentralized communication operates on a peer-to-peer basis among agents. Centralized communication involves a central agent or a group of central agents coordinating the system's communication, with other agents primarily connecting to the central agent. A shared message pool, as proposed in MetaGPT \cite{hong2023metagpt}, maintains a shared message pool where agents publish and subscribe to relevant messages, boosting communication efficiency.

\subsection{LLM-based Agents Management}
Research on the management of LLM-based agents is limited. Popular LLM-based multi-agent systems, such as MetaGPT \cite{hong2023metagpt}, AgentVerse \cite{chen2023agentverse}, and AutoGen \cite{wu2023autogen}, typically divide tasks into smaller sub-tasks and allocate multiple agents to complete them. However, their approaches to planning are sequential, lacking strategic management. In contrast, AIOS \cite{mei2024llm} introduces an LLM agent operating system that provides module isolation and integrates LLM and OS functions. It employs various managers, including Agent Scheduler, Context Manager, Memory Manager, Storage Manager, Tool Manager, and Access Manager, to effectively handle numerous agents. However, AIOS manually organizes different applications, such as a math problem-solving agent and a travel planning agent, rather than multiple agents within the same application. This approach represents a different type of SOP and is not applicable to large-scale MAS, as it is impractical for humans to write every SOP and prompt for each agent when the scale reaches thousands or even millions.

\subsection{Hallucinations in MAS}
Hallucination refers to the phenomenon where a model generates factually incorrect text \cite{zhao2023survey, huang2023survey}. Hallucinations are considered inevitable in LLMs \cite{banerjee2024llms}. This issue becomes more severe in MAS due to the multi-agent nature: one agent can send information to others. If an agent generates a hallucinated message, it may propagate to other agents, causing a cascading effect \cite{lee2024prompt, ju2024flooding}. Self-refinement through feedback and reasoning has proven effective, such as using self-reflection and prompting the LLM again to verify its outputs \cite{ji2023towards, tonmoy2024comprehensive}. Inspired by this, we equip \textit{MegaAgent} with a self-correction mechanism, enabling agents to review their outputs based on a to-do list generated at initialization. To enhance monitoring efficiency, we introduce a hierarchical monitoring mechanism: first, agents check their own outputs; second, an admin agent reviews the group's outputs; and third, a boss agent oversees the outputs of all groups.

\subsection{Novelty Comparison between MegaAgent and Baselines}
To highlight the distinctions between \textit{MegaAgent} and baseline models, we compare their supported features in \autoref{tab:relate-work}. The comparison shows that \textit{MegaAgent} stands out as the only MAS supporting key features, including: (1) No Pre-defined Standard Operating Procedures (SOPs); (2) Multi-file Input/Output Support; (3) Parallel Execution Capabilities; and (4) Scalability to a Large Number of Agents.

\begin{table*}[!ht]
\centering
\resizebox{\textwidth}{!}{%
\begin{tabular}{c|c|c|c|c|c}
\toprule
\rowcolor{champagne}\textbf{Feature}                    & \textbf{AutoGen} & \textbf{MetaGPT} & \textbf{CAMEL} & \textbf{AgentVerse} & \textbf{MegaAgent} \\ 
\midrule

\textbf{Definition of Each Agent's Task}          
& \makecell[c]{Users pre-define roles,\\ such as product manager \\and software engineer}   
& \makecell[c]{Users pre-define roles,\\ such as product manager \\and software engineer}     
& \makecell[c]{Pre-defined\\ agent abilities}        
& \makecell[c]{No Pre-defined\\ agent abilities}     
& \makecell[c]{No pre-defined \\ agent abilities}         \\ 

\midrule

\textbf{Support for Multi-File Input/Output}         
& \makecell[c]{Cannot handle\\ multiple files}        
& \makecell[c]{Can generate and manage\\ multiple files simultaneously}                             
& \makecell[c]{Cannot handle\\ multiple files}     
& \makecell[c]{Cannot handle\\ multiple files}     
& \makecell[c]{Can generate and manage\\ multiple files simultaneously}          \\ 
\midrule
\textbf{Support for Parallel Execution}                
& \makecell[c]{Tasks are finished sequentially,\\ one after another}        
& \makecell[c]{Tasks are finished sequentially,\\ one after another}        
& \makecell[c]{Tasks are completed sequentially,\\ one after another}   
& \makecell[c]{Tasks are completed sequentially,\\ one after another} 
& \makecell[c]{Tasks are completed in parallel}           \\ 
\midrule
\textbf{Scalability to Large Numbers of Agents}                
& \makecell[c]{Restricted by the number\\ of user-defined agents}     
& \makecell[c]{Limited by the number of\\ user-defined agents}        
& \makecell[c]{Limited by the number of\\ user-defined agents}   
& \makecell[c]{Limited by the number of\\ user-defined agents}   
& \makecell[c]{Can adaptively generate more agents\\ based on needs of the task}         \\ 

\bottomrule
\end{tabular}
}
\caption{Comparison of features across MAS. 
\textbf{Definition of Each Agent's Task}: Indicates whether the system can autonomously produce a clear and customizable definition of roles and tasks for individual agents. Both MegaAgent and AgentVerse support this feature, while other systems rely on fixed or developer-specified tasks. 
\textbf{Support for Multi-File Input/Output}: Refers to the ability of systems to process and manage multiple files simultaneously. MegaAgent and MetaGPT support this functionality, enhancing their usability for complex workflows. 
\textbf{Support for Parallel Execution}: Indicates whether the system can execute multiple tasks in parallel. Only MegaAgent supports true parallel execution, while other systems operate sequentially.
\textbf{Scalability to Large Numbers of Agents}: Assesses the system’s capability to scale efficiently when the number of agents increases. MegaAgent is the only system designed to handle a large number of agents seamlessly, demonstrating superior scalability.}
\label{tab:relate-work}
\end{table*}

\section{Travel Planner Experiment}

We further evaluate MegaAgent on the validation set of TravelPlanner \cite{xie2024travelplanner}, a benchmark designed for real-world planning with multiple constraints. The environment contains over four million data records accessible via six tools, and includes 1,225 planning intents with corresponding reference plans.

\subsection{Task Description}

Each planning task involves a variety of constraints:
\begin{itemize}
    \item \textbf{Environment Constraints}: Real-world limitations such as unavailable flights.
    \item \textbf{Commonsense Constraints}: Reasonable expectations, e.g., avoiding repeated attractions.
    \item \textbf{Hard Constraints}: User-specific requirements, such as budget, cuisine, and lodging rules.
\end{itemize}

\subsection{Results}

MegaAgent is evaluated in the \emph{sole-planning} mode, where destination cities and necessary information (e.g.\ restaurants) are provided to the agent, eliminating tool calls and focusing purely on planning. Results on the validation set (\#180) are shown in Table~\ref{tab:travelplanner_sole}.

\begin{table*}[!ht]
\centering
\resizebox{\textwidth}{!}{%
\begin{tabular}{l|cccccc}
\toprule
\rowcolor{champagne}\textbf{Planning Strategy}       & \textbf{DR (\%)} & \textbf{CS-Micro (\%)} & \textbf{CS-Macro (\%)} & \textbf{HC-Micro (\%)} & \textbf{HC-Macro (\%)} & \textbf{FPR (\%)} \\
\midrule
\rowcolor{gray!10}Direct GPT-3.5-Turbo \cite{openai2022}            & 100.0            & 60.2                   & 4.4                    & 11.0                   & 2.8                    & 0.0               \\
CoT GPT-3.5-Turbo \cite{wei2023chainofthoughtpromptingelicitsreasoning}             & 100.0            & 66.3                   & 3.3                    & 11.9                   & 5.0                    & 0.0               \\
\rowcolor{gray!10}ReAct GPT-3.5-Turbo \cite{yao2022react}              &  82.2            & 47.6                   & 3.9                    & 11.4                   & 6.7                    & 0.6               \\
Reflexion GPT-3.5-Turbo \cite{shinn2023reflexion}    &  93.9            & 53.8                   & 2.8                    & 11.0                   & 2.8                    & 0.0               \\
\rowcolor{gray!10}Direct Mixtral-8×7B-MoE \cite{jiang2024mixtralexperts} & 100.0          & 68.1                   & 5.0                    &  3.3                   & 1.1                    & 0.0               \\
Direct Gemini Pro \cite{geminiteam2025geminifamilyhighlycapable} & 93.9   & 65.0                   & 8.3                    &  9.3                   & 4.4                    & 0.6               \\
\rowcolor{gray!10}Direct GPT-4-Turbo \cite{openai2023gpt}  & 100.0          & 80.4                   & 17.2                   & \textbf{47.1}                   & 22.2                   & 4.4               \\
\midrule
\textbf{MegaAgent}                                   & \textbf{100.0}   & \textbf{81.88}         & \textbf{27.22}         & 40.48         & \textbf{23.89}         & \textbf{10.0}     \\
\bottomrule
\end{tabular}%
}
\caption{Comparison of MegaAgent and TravelPlanner planning strategies on the sole-planning mode (Validation Set, \#180).}
\label{tab:travelplanner_sole}
\end{table*}

\section{Gobang Game Experiment Details} \label{megaagent-gobang}

\subsection{Setup}
We use ChatGPT-4o API for this experiment. The 'temperature' parameter is set to 0 to reduce the randomness of the outputs \cite{achiam2023gpt}.

\subsection{Cost}
The total cost comparision is shown in \autoref{tab:GobangCost}. Despite using more tokens, MegaAgent delivers superior results at reasonable costs. For the Gobang experiment, while MegaAgent's cost (\$6.90) is higher than some baselines, it is the only system that successfully completes the task.

\begin{table}[h]
\centering
\begin{tabular}{lrrr}
\toprule
\rowcolor{champagne}\textbf{Model} & \textbf{Cost (USD)} & \textbf{\# Agents} & \textbf{Time (s)} \\
\midrule
\rowcolor{gray!10}AutoGen & 0.10 & 2 & 180 \\
MetaGPT & 0.61 & 6 & 480 \\
\rowcolor{gray!10}CAMEL & 0.76 & 2 & 1,830 \\
AgentVerse & 8.07 & 4 & 1,980 \\
\rowcolor{gray!10}\textbf{MegaAgent} & \textbf{6.90} & \textbf{7} & \textbf{800} \\
\bottomrule
\end{tabular}
\caption{Cost Comparison for Gobang Game Development}
\label{tab:GobangCost}
\end{table}

\subsection{Results}
First, Boss Agent receives the initial hand-written meta-prompt, shown in \autoref{fig:gobangPrompt}. Then, \textit{MegaAgent} utilizes these initial prompts as the system message to create agents, with additional written function calls in \autoref{fig:function_calls_part1} and \autoref{fig:function_calls_part2}. The communication content and function call results are added directly into the corresponding agent's memory. Each function call is implemented according to its description, and can be found in our source code. The initial prompt and the additional written functions are the only prompts that are written by hand, showcasing our framework's autonomy.

\subsection{Ablation Study} \label{gobang-ablation}
We conduct the ablation study of \textit{MegaAgent} for the Gobang task. We rerun the experiment without hierarchy, parallelism, and monitoring mechanism, separately. 

\begin{figure}[htbp]
    \centering
    \includegraphics[width=0.95\linewidth]{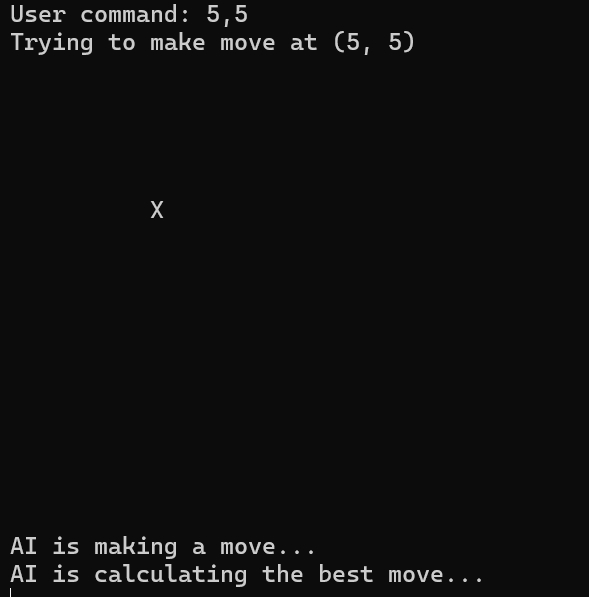}
    \caption{Failure of MegaAgent without Hierarchy}
    \label{fig:infinite_loop}
\end{figure}

When running without hierarchy, group managers cannot create new agents. As shown in \autoref{fig:infinite_loop}, the generated program will fall in an infinite loop. However, the AI development group's manager cannot resolve this issue by himself. Nor can he recruit new agents for collaboration in this scenario.

When running without parallelism, each group will complete their tasks one by one, linearly. Although this will not hinder the system's performance, the time complexity will drop from $O(\log n)$ to $O(n)$. As a result, the execution time grows from  800 seconds to 4505 seconds.

\begin{figure}[htbp]
    \centering
    \includegraphics[width=0.95\linewidth]{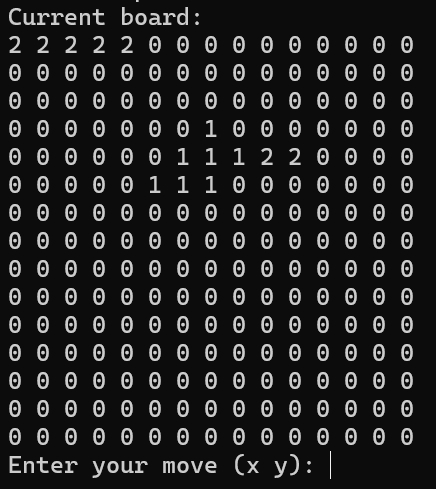}
    \caption{Failure of MegaAgent without the Monitoring Mechanism}
    \label{fig:not_terminate}
\end{figure}

When running without the monitoring mechanism, the group leaders will not validate the program. As shown in figure \autoref{fig:not_terminate}, the program cannot terminate when there are five-in-a-row, but the group agents do not find this bug because of the lack of the monitoring mechanism.

\begin{figure}[htp]
\centering
\begin{tikzpicture}
    \definecolor{chatcolor1}{HTML}{FFFACD}
    \definecolor{shadecolor}{gray}{0.95}

    % Define the first node
    \node[
        align=left,
        text width=0.45\textwidth,
        fill=shadecolor,
        rounded corners=1mm,
        draw=black,
        line width=0.5mm
    ] (node1) at (0,0) {
        \textit{You are Bob, the leader of a software development club. Your club's current goal is to develop a Gobang game with a very strong AI, no frontend, and can be executed by running \texttt{'main.py'}. You are now recruiting employees and assigning work to them. For each employee (including yourself), please write a prompt specifying: their name (one word, no prefix), their job, the tasks they need to complete, and their collaborators' names and jobs. The format should follow the example below:}
        
        \texttt{<employee name="Alice">\\
        You are Alice, a novelist. Your job is to write a single chapter of a novel with 1000 words according to the outline (outline.txt) from Carol, the architect designer, and pass it to David (chapter\_x.txt), the editor. Please only follow this routine. Your collaborators include Bob (the Boss), Carol (the architect designer), and David (the editor).\\
        </employee>}
        
        \textit{Please note that every employee is lazy and will only perform the tasks explicitly mentioned in their prompt. To ensure project completion, each task must be non-divisible, detailed, specific, and involve only supported file types (txt or python). You should recruit enough employees to cover the entire SOP, ensuring tasks are distributed to speed up the process. Finally, specify an employee's name to initiate the project in the format:}
        
        \texttt{<beginner>Name</beginner>}
    };

\end{tikzpicture}
\caption{Gobang Game Development Meta Prompt}
\label{fig:gobangPrompt}
\end{figure}

Then, \textit{MegaAgent} would generate different agent roles in \autoref{fig:gobang_roles}. After generation, each agent will update its own TODO list, utilize function calls to complete its tasks, or talk to other agents, until it clears its TODO list and marks its task as 'Done'. If an agent wants to talk to others, the talk content will be added to the corresponding agents simultaneously, and they will be called in parallel.

\begin{figure}[htp]
\centering
\begin{tcolorbox}[
    colback=orange!30,    % Deeper orange background for better visibility
    colframe=orange!80,   % Stronger orange frame for clear contrast
    boxrule=0.5mm, 
    rounded corners, 
    width=\linewidth, 
  title=Function Calls for Gobang Game Development (Part 1)
]

\scriptsize
\texttt{
\{"name": "exec\_python\_file", \\
\quad "description": "Execute a Python file and get the result.", \\
\quad "parameters": \{ \\
\quad\quad "type": "object", \\
\quad\quad "properties": \{ \\
\quad\quad\quad "filename": \{ \\
\quad\quad\quad\quad "type": "string", \\
\quad\quad\quad\quad "description": "The filename of the Python file to be executed." \\
\quad\quad\quad \} \\
\quad\quad \} \\
\quad \} \\
\}, \\
\{"name": "read\_file", \\
\quad "description": "Read the content of a file.", \\
\quad "parameters": \{ \\
\quad\quad "type": "object", \\
\quad\quad "properties": \{ \\
\quad\quad\quad "filename": \{ \\
\quad\quad\quad\quad "type": "string", \\
\quad\quad\quad\quad "description": "The filename to be read." \\
\quad\quad\quad \} \\
\quad\quad \} \\
\quad \} \\
\}, \\
\{"name": "input", \\
\quad "description": "Input a string to the running Python code.", \\
\quad "parameters": \{ \\
\quad\quad "type": "object", \\
\quad\quad "properties": \{ \\
\quad\quad\quad "content": \{ \\
\quad\quad\quad\quad "type": "string", \\
\quad\quad\quad\quad "description": "The string to be input." \\
\quad\quad\quad \} \\
\quad\quad \} \\
\quad \} \\
\} \\
}
\end{tcolorbox}
\caption{Function Calls for Gobang Game Development (Part 1).}
\label{fig:function_calls_part1}
\end{figure}

\begin{figure}[htp]
\centering
\begin{tcolorbox}[
    colback=orange!30,    % Deeper orange background for better visibility
    colframe=orange!80,   % Stronger orange frame for clear contrast
    boxrule=0.5mm, 
    rounded corners, 
    width=\linewidth, 
  title=Function Calls for Gobang Game Development (Part 2)
]

\scriptsize
\texttt{
\{"name": "write\_file", \\
\quad "description": "Write content to a file.", \\
\quad "parameters": \{ \\
\quad\quad "type": "object", \\
\quad\quad "properties": \{ \\
\quad\quad\quad "filename": \{ \\
\quad\quad\quad\quad "type": "string", \\
\quad\quad\quad\quad "description": "The filename to be written." \\
\quad\quad\quad \}, \\
\quad\quad\quad "content": \{ \\
\quad\quad\quad\quad "type": "string", \\
\quad\quad\quad\quad "description": "The content to be written." \\
\quad\quad\quad \} \\
\quad\quad \} \\
\quad \} \\
\}, \\
\{"name": "add\_agent", \\
\quad "description": "Recruit an agent as your subordinate.", \\
\quad "parameters": \{ \\
\quad\quad "type": "object", \\
\quad\quad "properties": \{ \\
\quad\quad\quad "name": \{ \\
\quad\quad\quad\quad "type": "string", \\
\quad\quad\quad\quad "description": "Unique agent name." \\
\quad\quad\quad \}, \\
\quad\quad\quad "description": \{ \\
\quad\quad\quad\quad "type": "string", \\
\quad\quad\quad\quad "description": "Agent description." \\
\quad\quad\quad \} \\
\quad\quad \} \\
\quad \} \\
\}, \\
\{"name": "TERMINATE", \\
\quad "description": "End the conversation when all tasks are complete." \\
\} \\
}
\end{tcolorbox}
\caption{Function Calls for Gobang Game Development (Part 2).}
\label{fig:function_calls_part2}
\end{figure}

\begin{figure}[htp]
\centering
\begin{tikzpicture}
    \definecolor{shadecolor}{gray}{0.95}

    % Bob's Role
    \node[
        align=left,
        text width=0.45\textwidth,
        fill=shadecolor,
        rounded corners=1mm,
        draw=black,
        line width=0.5mm
    ] (node1) at (0,0) {
        \textit{You are Bob, the leader of the software development club. Your job is to decide all the features to develop for the Gobang game and write them in a file named \texttt{'features.txt'}. Your collaborators include Alice (game designer), Carol (AI developer), David (game logic developer), and Eve (integrator).}
    };

    % Alice's Role
    \node[
        align=left,
        text width=0.45\textwidth,
        fill=shadecolor,
        rounded corners=1mm,
        draw=black,
        line width=0.5mm
    ] (node2) at (0,-4) {
        \textit{You are Alice, a game designer. Your job is to design the game rules and user interactions based on the features listed in \texttt{'features.txt'} from Bob, and document them in a file named \texttt{'game\_design.txt'}. Your collaborators include Bob (leader), Carol (AI developer), David (game logic developer), and Eve (integrator).}
    };

    % Carol's Role
    \node[
        align=left,
        text width=0.45\textwidth,
        fill=shadecolor,
        rounded corners=1mm,
        draw=black,
        line width=0.5mm
    ] (node3) at (0,-8) {
        \textit{You are Carol, an AI developer. Your job is to develop the AI for the Gobang game based on the game design in \texttt{'game\_design.txt'} from Alice, and write the AI code in a file named \texttt{'ai.py'}. Your collaborators include Bob (leader), Alice (game designer), David (game logic developer), and Eve (integrator).}
    };

    % David's Role
    \node[
        align=left,
        text width=0.45\textwidth,
        fill=shadecolor,
        rounded corners=1mm,
        draw=black,
        line width=0.5mm
    ] (node4) at (0,-12) {
        \textit{You are David, a game logic developer. Your job is to develop the game logic for the Gobang game based on the game design in \texttt{'game\_design.txt'} from Alice, and write the game logic code in a file named \texttt{'game\_logic.py'}. Your collaborators include Bob (leader), Alice (game designer), Carol (AI developer), and Eve (integrator).}
    };

    % Eve's Role
    \node[
        align=left,
        text width=0.45\textwidth,
        fill=shadecolor,
        rounded corners=1mm,
        draw=black,
        line width=0.5mm
    ] (node5) at (0,-16.5) {
        \textit{You are Eve, an integrator. Your job is to integrate the AI code from \texttt{'ai.py'} by Carol and the game logic code from \texttt{'game\_logic.py'} by David, and write the integration code in a file named \texttt{'main.py'} to ensure the Gobang game can be executed by running \texttt{'main.py'}. Your collaborators include Bob (leader), Alice (game designer), Carol (AI developer), and David (game logic developer).}
    };

\end{tikzpicture}
\caption{Role Assignments Generated by MegaAgent}
\label{fig:gobang_roles}
\end{figure}

The memory of each agent is implemented by a chroma vector database \footnote{https://www.trychroma.com/}. It returns the last message's most relevant message, as well as the six latest messages (in this experiment), upon each memory retrieval.

\begin{figure}[htbp]
    \centering
    \includegraphics[width=0.95\linewidth]{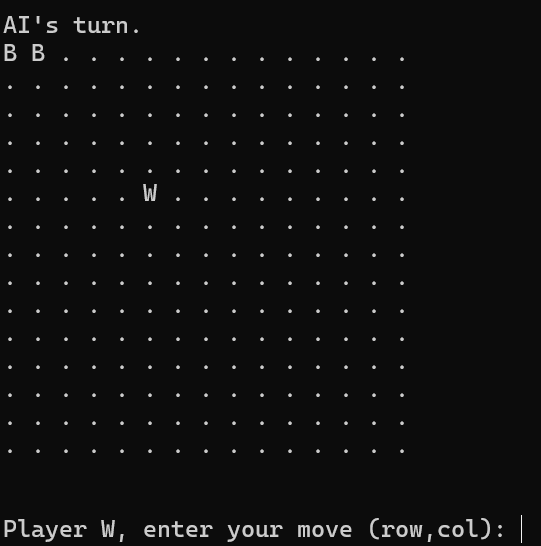}
    \caption{Interface of Gobang demo produced by MegaAgent}
    \label{fig:gobangdemo}
\end{figure}

In our experiment, \textit{MegaAgent} successfully produces a runnable Gobang game with a naive AI upon the first trial, whose interface is shown in \autoref{fig:gobangdemo}. 

\subsection{Human-written SOP for Gobang Game}
To evaluate the performance of the Gobang Game development against other baselines, we provide a \textbf{human-written SOP} for the Gobang Game, as shown in \autoref{fig:human-written-gobang}. This serves as a benchmark for comparison with the \textit{MegaAgent}-generated SOP.

\begin{figure}[htp]
\centering
\begin{tikzpicture}
    \definecolor{shadecolor}{gray}{0.95}

    % Vertical spacing adjustment
    \def\verticalSpacing{3.5}

    % Bob's Role
    \node[
        align=left,
        text width=0.45\textwidth,
        fill=shadecolor,
        rounded corners=1mm,
        draw=black,
        line width=0.5mm
    ] (node1) at (0, 0.5) {
        \textit{You are Bob, the boss of the software development team. You are responsible for monitoring the project's progress and ensuring that it can be executed by running the \texttt{main.py} file in the end. Your team members are Alan (game logic design), Alice (\texttt{board.py}), Charlie (\texttt{main.py}), David (\texttt{ai.py}), and Emily (testing).}
    };

    % Alan's Role
    \node[
        align=left,
        text width=0.45\textwidth,
        fill=shadecolor,
        rounded corners=1mm,
        draw=black,
        line width=0.5mm
    ] (node2) at (0, -\verticalSpacing) {
        \textit{You are Alan, an architect designer. Your job is to design the game logic of the Gobang game and propose possible AI implementations. Document your design in \texttt{design.txt} and pass it to your teammates. Collaborators: Bob (Boss), Alice (\texttt{board.py}), Charlie (\texttt{main.py}), David (\texttt{ai.py}).}
    };

    % Alice's Role
    \node[
        align=left,
        text width=0.45\textwidth,
        fill=shadecolor,
        rounded corners=1mm,
        draw=black,
        line width=0.5mm
    ] (node3) at (0, -2*\verticalSpacing) {
        \textit{You are Alice, a software developer. Implement the \texttt{board.py} file based on Alan's design in \texttt{design.txt}. Collaborators: Bob (Boss), Alan (game logic), Charlie (\texttt{main.py}), David (\texttt{ai.py}), Emily (testing).}
    };

    % Charlie's Role
    \node[
        align=left,
        text width=0.45\textwidth,
        fill=shadecolor,
        rounded corners=1mm,
        draw=black,
        line width=0.5mm
    ] (node4) at (0, -3*\verticalSpacing) {
        \textit{You are Charlie, a software developer. Implement the \texttt{main.py} file based on Alan's design in \texttt{design.txt}. Ensure compatibility with \texttt{board.py} (Alice) and \texttt{ai.py} (David). Optionally create \texttt{test.py} for testing. Collaborators: Bob (Boss), Alan (game logic), Alice (\texttt{board.py}), David (\texttt{ai.py}), Emily (testing).}
    };

    % David's Role
    \node[
        align=left,
        text width=0.45\textwidth,
        fill=shadecolor,
        rounded corners=1mm,
        draw=black,
        line width=0.5mm
    ] (node5) at (0, -4*\verticalSpacing) {
        \textit{You are David, an AI developer. Implement a naive \texttt{ai.py} file that makes random moves quickly. Collaborators: Bob (Boss), Alan (game logic), Alice (\texttt{board.py}), Charlie (\texttt{main.py}), Emily (testing).}
    };

    % Emily's Role
    \node[
        align=left,
        text width=0.45\textwidth,
        fill=shadecolor,
        rounded corners=1mm,
        draw=black,
        line width=0.5mm
    ] (node6) at (0, -5*\verticalSpacing) {
        \textit{You are Emily, a tester. Test the Gobang game's correctness and efficiency. Write \texttt{test.py} and ensure the game runs correctly by executing \texttt{main.py}. Test thoroughly until the game completes. Collaborators: Bob (Boss), Alan (game logic), Alice (\texttt{board.py}), Charlie (\texttt{main.py}), David (\texttt{ai.py}).}
    };

\end{tikzpicture}
\caption{Human-written Prompts for Gobang Game Development}
\label{fig:human-written-gobang}
\end{figure}

\subsection{Gobang Game Experiment with Baselines} \label{sec-gobangbaseline}
We conduct the same Gobang game task experiment on state-of-the-art MAS as of July 2024.

\subsubsection{AutoGen Setup and Result} \label{autogen-gobang}
We test AutoGen v1.0.16 based on its multi-agent coding demo. We only fill in the API key and change its prompt to: \textit{Develop a Gobang game with an AI} , and leave everything else unchanged. We do not allow runtime human input.

\begin{figure*}[htbp]
    \centering
    \includegraphics[width=0.95\linewidth]{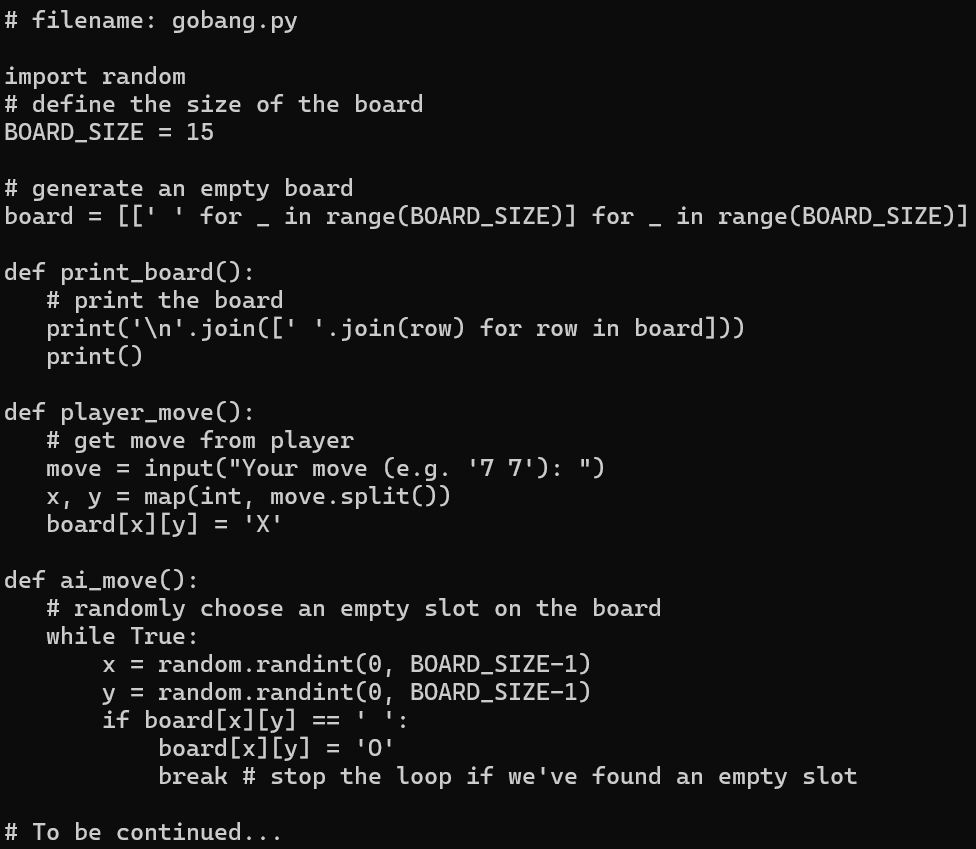}
    \caption{Code Produced by AutoGen}
    \label{fig:autogen1}
\end{figure*}

\begin{figure*}[htbp]
    \centering
    \includegraphics[width=0.95\linewidth]{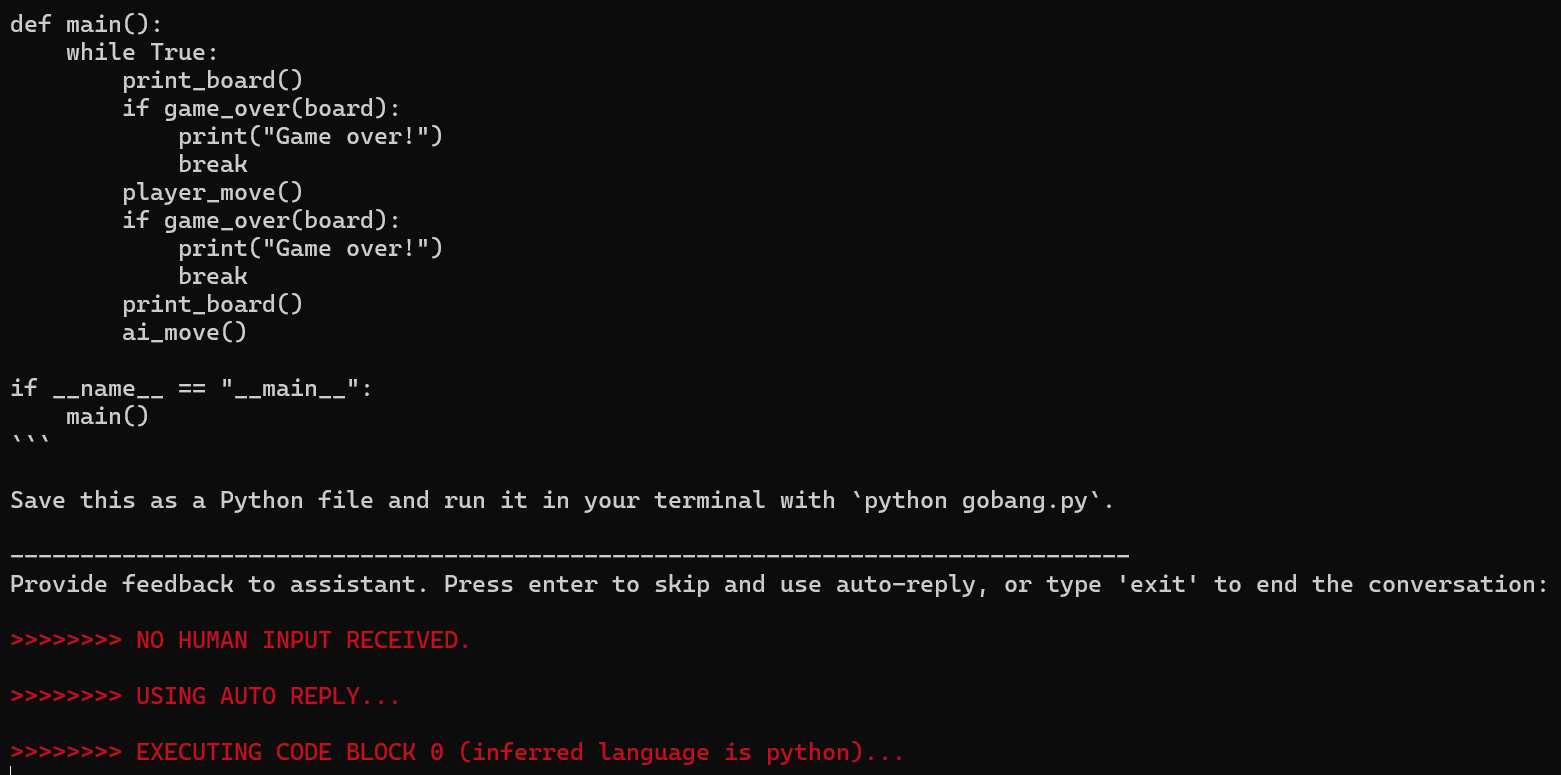}
    \caption{Execution Result of AutoGen}
    \label{fig:autogen2}
\end{figure*}

As shown in \autoref{fig:autogen1} and \autoref{fig:autogen2}, AutoGen generates a program ending with \texttt{\# To be continued..} after about two minutes, and gets stuck when trying to execute it. The possible reason for its failure is that its SOP is too simple and does not include enough communication e.g. code review between agents.

\begin{figure*}
    \centering
    \includegraphics[width=0.95\linewidth]{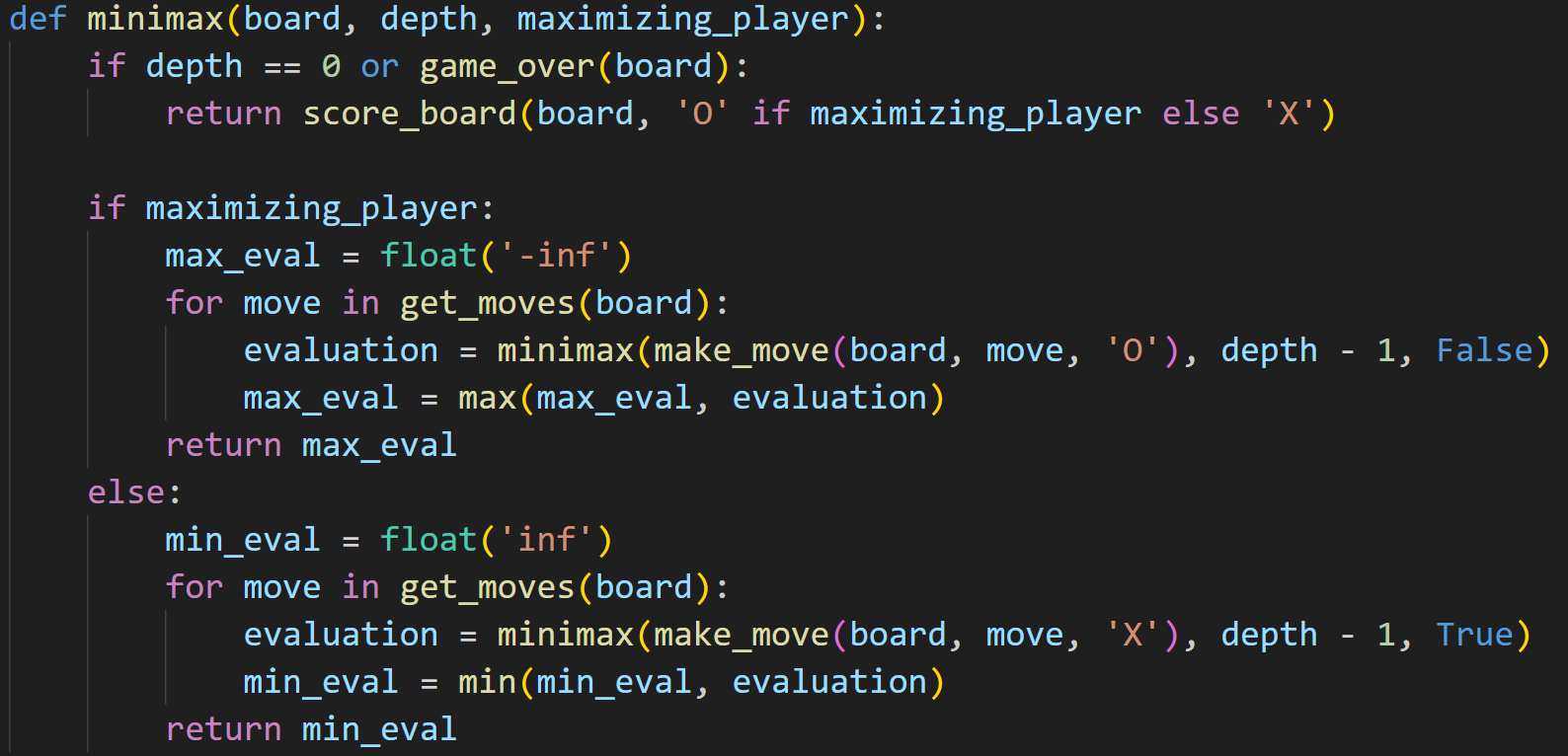}
    \caption{Code Produced by AutoGen in Another Trial}
    \label{fig:autogen4}
\end{figure*}

\begin{figure*}
    \centering
    \includegraphics[width=0.95\linewidth]{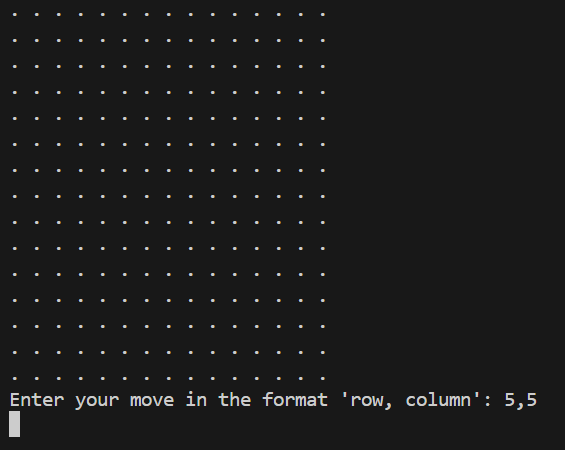}
    \caption{Execution Result of AutoGen in Another Trial. AI will keep thinking for almost infinite time. }
    \label{fig:autogen3}
\end{figure*}

We try three times, which all end with similar results. In another one trial, as shown in Figure \ref{fig:autogen4} and Figure \ref{fig:autogen3}, AutoGen successfully produces an AI with mini-max algorithm, but no pruning. This is impossible to execute in a limited time, as the state space of Gobang game is very large. We try another prompt: \textit{Develop a Gobang game with a very strong AI, no frontend, and can be executed by
running 'main.py'} , and get similar results.

By the time it gets stuck, AutoGen has cost \$0.1 and 120 seconds. Since AutoGen cannot complete this task, we are unable to count the overall cost.

\subsubsection{MetaGPT Setup and Result} \label{metagpt-gobang}
\begin{figure*}[htbp]
    \centering
    \includegraphics[width=0.95\linewidth]{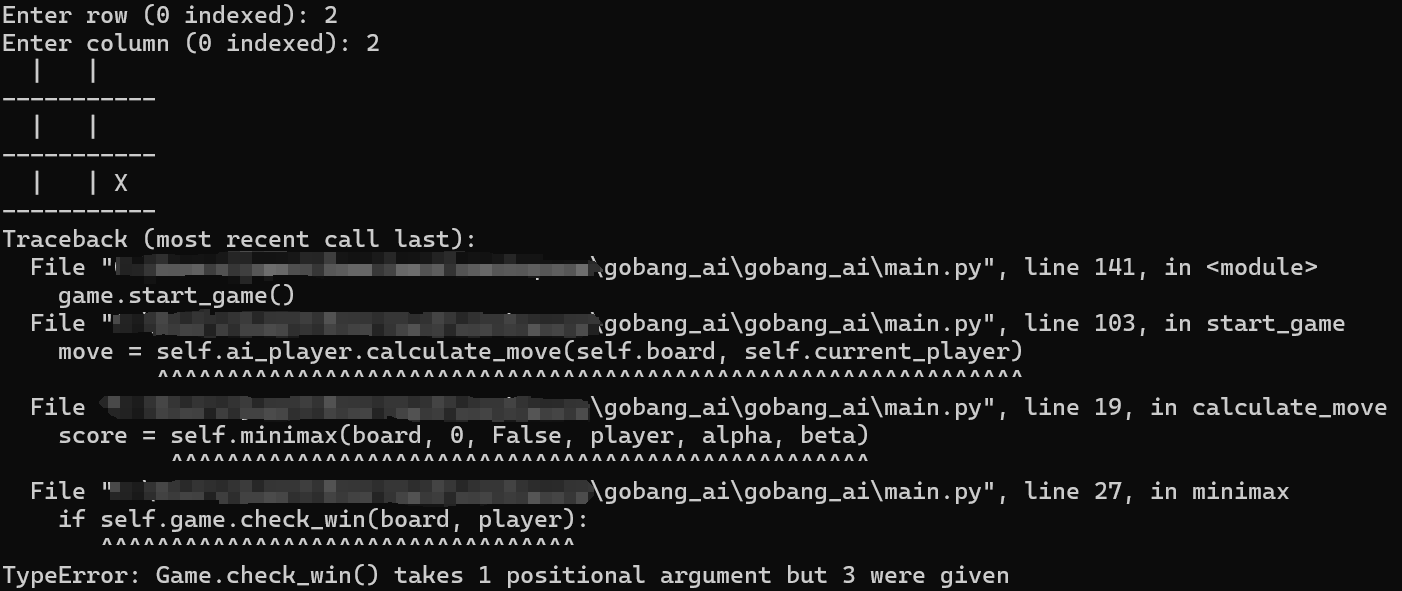}
    \caption{Execution Result of the Code Produced by MetaGPT}
    \label{fig:metagpt}
\end{figure*}

We test MetaGPT v0.8.1 by feeding the prompt: \textit{Develop a Gobang game with an AI}. We fill in the API key and leave everything else unchanged. It produces results in \autoref{fig:metagpt}, and its execution time is around eight minutes. We try three times, and find none of them can produce an AI move. The major errors are:
\begin{itemize}
    \item \textbf{The code is not executable, and raises an error.} The possible reason is that MetaGPT does not have external tools to execute and debug the produced code.
    \item \textbf{The produced program is not a Gobang game (for example, a tic-tac-toe game instead).} The possible reason for failure is that its SOP is too simple, and the requirement for communication between agents is not sufficient.
    \item \textbf{AI falls into an infinite loop.} The possible reason is that MetaGPT does not have external tools to execute and debug the produced code, and the current ChatGPT API is not capable of developing the AlphaBeta algorithm without errors by itself.
\end{itemize}

\subsubsection{CAMEL Setup and Result} \label{camel-gobang}
\begin{figure*}
    \centering
    \includegraphics[width=0.95\linewidth]{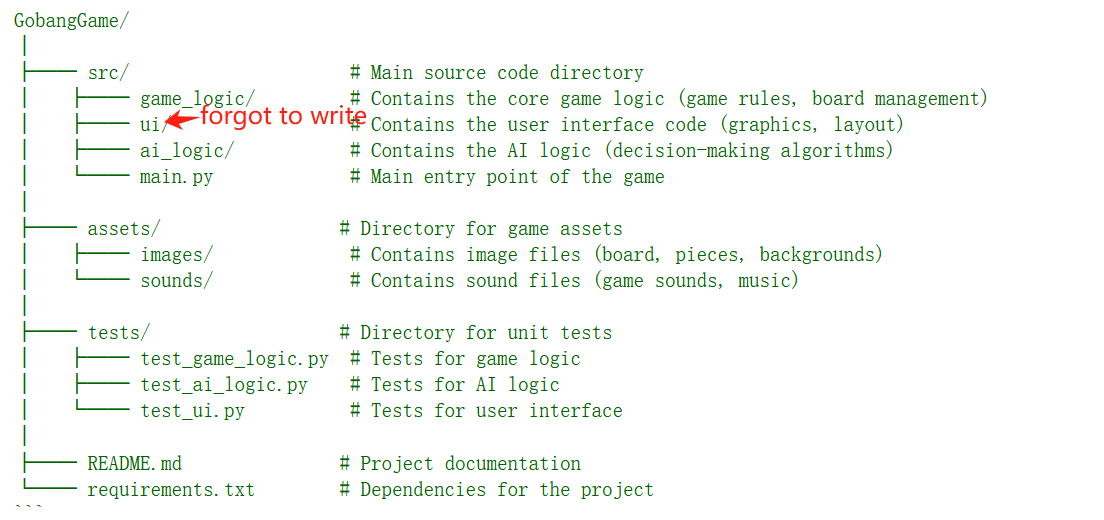}
    \caption{An example of CAMEL's output. It forgets to write ui.py in this trial.}
    \label{fig:camel}
\end{figure*}

We use the CAMEL v0.1.6.0 Jupiter Notebook demo in Colab. We fill in the API key, change the task prompt to: \textit{Develop a Gobang game with an AI}, and leave everything else unchanged. We try three times. It turns out that CAMEL can only produce code segments. For example, in one trial, as shown in \autoref{fig:camel}, CAMEL forgets to write \texttt{ui.py}, which is included in \texttt{game.py}. The possible reason for this is that its planning and contextual ability are weak. The total cost of one trial is \$0.76.

\subsubsection{AgentVerse Setup and Result} \label{agentverse-gobang}
\begin{figure*}
    \centering
    \includegraphics[width=0.95\linewidth]{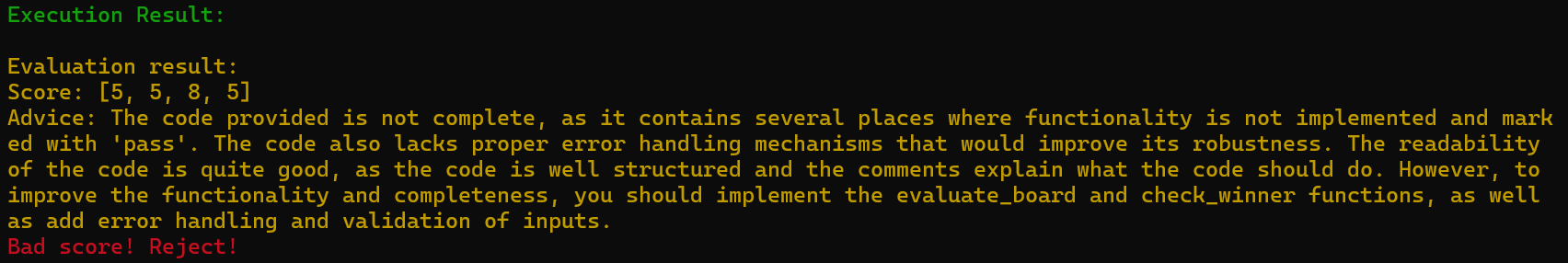}
    \caption{A rejected trial of AgentVerse after ten rounds. }
    \label{fig:agentverse}
\end{figure*}

\begin{figure*}
    \centering
    \includegraphics[width=0.95\linewidth]{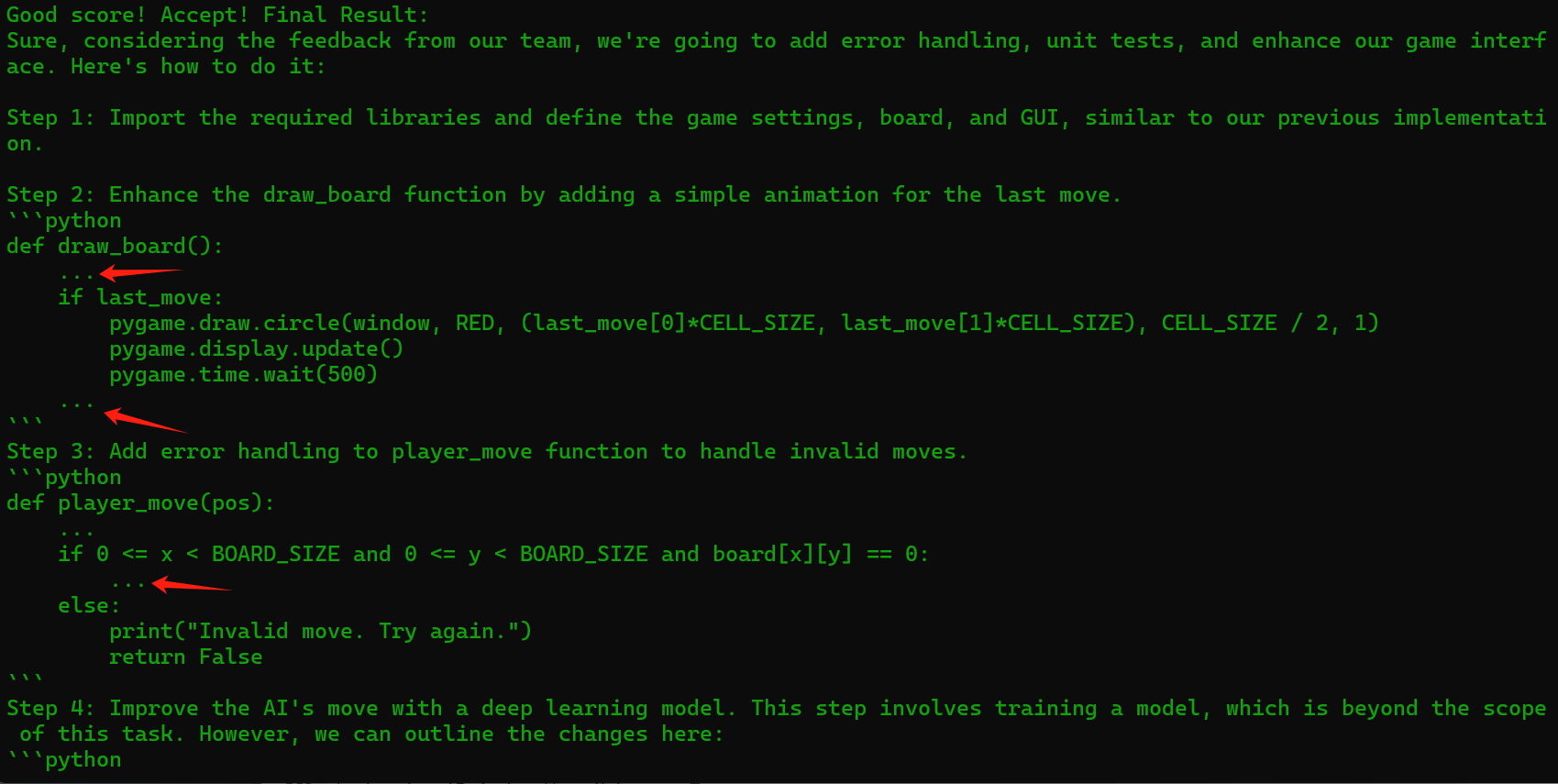}
    \caption{An accepted trial of AgentVerse. The code still contains many placeholders, and is not executable.}
    \label{fig:agentverse2}
\end{figure*}

We test AgentVerse v0.1.8.1 based on its \texttt{tasksolving/pythoncalculator} scenario. We fill in the API key, change the \texttt{max\_turn} parameter from 3 to 10 to allow more rounds for better results, and modify the task description to: \textit{develop a Gobang game with an AI using Python3.} We leave everything else unchanged and try three times. We find that, in the first and second trial, the agent keeps rejecting the result for all the ten rounds, as shown in \autoref{fig:agentverse}; as for the third trial, although the agent accepts the result, the code as shown in \autoref{fig:agentverse2} still presents many placeholders, and cannot be executed. Given that ten rounds significantly exceed the default setting, we conclude that AgentVerse is unlikely to successfully complete the Gobang task even with additional rounds and opportunities. One trial costs about \$8.07, and 1980 seconds.

To sum up, our \textit{MegaAgent} framework is the \textbf{first and only} MAS to develop the Gobang game successfully.

\subsection{Comparison of the token usage}

We provide a rough analysis of the token distribution for each baseline. The results are shown in \autoref{tab:gobang_cost}.
\begin{table*}[ht]
\centering
\resizebox{0.8\textwidth}{!}{%
\begin{tabular}{c|c|ccc|c}
\toprule
\rowcolor{champagne}\textbf{Model} & \textbf{Stage} & \textbf{Input Tokens} & \textbf{Output Tokens} & \textbf{Total Tokens} & \textbf{Time (s)} \\
\midrule
\rowcolor{gray!10}
& Planning     & 0      & 0      & 0      & N/A     \\
\rowcolor{gray!10} & Task-solving & 93     & 845    & 938    & 0–40    \\
\rowcolor{gray!10} & Merging      & 0      & 0      & 0      & N/A     \\
\rowcolor{gray!10} \multirow{-4}{*}{AutoGen} & Total        & 93     & 845    & 938    & 40      \\
\midrule
\multirow{4}{*}{MetaGPT} 
& Planning     & 1,488  & 612    & 2,100  & 0–107   \\
& Task-solving & 62,562 & 12,550 & 75,112 & 107–990 \\
& Merging      & 5,981  & 43     & 6,024  & 990–1,034 \\
& Total        & 70,031 & 13,205 & 83,236 & 1,034   \\
\midrule
\rowcolor{gray!10}
& Planning     & 116    & 633    & 749    & 0–15    \\
\rowcolor{gray!10} & Task-solving & 80,985 & 15,741 & 96,726 & 10–1,830 \\
\rowcolor{gray!10} & Merging      & 0      & 0      & 0      & N/A     \\
\rowcolor{gray!10} \multirow{-4}{*}{CAMEL} & Total        & 81,101 & 16,374 & 97,475 & 1,830   \\
\midrule
\multirow{4}{*}{AgentVerse} 
& Planning     & 3,200  & 1,204  & 4,404  & 0–70    \\
& Task-solving & 179,631 & 44,685 & 224,316 & 70–1,282 \\
& Merging      & 0      & 0      & 0      & N/A     \\
& Total        & 182,831 & 45,889 & 228,720 & 1,282   \\
\midrule
\rowcolor{gray!10}
& Planning     & 42,947 & 12,347 & 55,294 & 0–60    \\
\rowcolor{gray!10} & Task-solving & 1,098,573 & 55,022 & 1,153,595 & 30–840 \\
\rowcolor{gray!10} & Merging      & 22,099 & 1,493  & 23,592 & 840–870 \\
\rowcolor{gray!10}\multirow{-4}{*}{MegaAgent}  & Total        & 1,163,619 & 68,862 & 1,232,481 & 870     \\
\bottomrule
\end{tabular}
}
\caption{Comparison of the estimated token and time distribution of all baselines in Gobang Game development }
\label{tab:gobang_cost}
\end{table*}

\section{National Policy Generation Experiment Details}  \label{megaagent-national}

\subsection{Setup}
We use the ChatGPT-4o mini API for this experiment. The `temperature' parameter is set to default. The memory of each agent returns the most relevant message, as well as ten latest messages in this experiment.

\subsection{Cost}
The total cost of this experiment is shown in \autoref{tab:PolicyCost}. Despite using more tokens, MegaAgent delivers superior results at reasonable costs. For the National Policy Generation, MegaAgent coordinates 590 agents (compared to 1-6 agents for baselines) while maintaining a competitive cost of \$3.30, demonstrating exceptional scalability.

\begin{table}[h]
\centering
\begin{tabular}{lrrr}
\toprule
\rowcolor{champagne}\textbf{Model} & \textbf{Cost (USD)} & \textbf{\# Agents} & \textbf{Time (s)} \\
\midrule
\rowcolor{gray!10}AutoGen & 0.10 & 1 & 40 \\
MetaGPT & 0.90 & 6 & 580 \\
\rowcolor{gray!10}CAMEL & 0.68 & 2 & 1,380 \\
AgentVerse & 2.05 & 4 & 510 \\
\rowcolor{gray!10}\textbf{MegaAgent} & \textbf{3.30} & \textbf{590} & \textbf{2,991} \\
\bottomrule
\end{tabular}
\caption{Cost Comparison for National Policy Generation}
\label{tab:PolicyCost}
\end{table}

\subsection{Results} \label{megaagent-result}
Boss agent receives the initial hand-written meta-prompt in \autoref{fig:nationleader_prompt}. Then, Boss Agent generated several admin agents shown in \autoref{fig:policy_roles}.

\begin{figure}[htbp]
\centering
\begin{tikzpicture}
    \definecolor{shadecolor}{gray}{0.95}

    % Full Prompt Content
    \node[
        align=left,
        text width=0.45\textwidth,
        fill=shadecolor,
        rounded corners=1mm,
        draw=black,
        line width=0.5mm
    ] (node1) at (0,0) {
        \textbf{NationLeader's Prompt:} \\
        
        You are NationLeader, the leader of a pioneering nation. You want to develop the best detailed policy for your cutting-edge country in \texttt{'policy\_\{department\}.txt'}. You are now recruiting ministers and assigning work to them. For each possible minister, please write a prompt. Please specify his name (no space), his job, and what kinds of work he needs to do. Note that each of them can recruit subordinates and conduct tests on them based on your policy. You \textbf{MUST} clarify all his possible collaborators' names and their jobs in the prompt. The format should be like (The example is for Alice in another novel writing project): 
        
        \vspace{0.3cm}
        \texttt{<employee name="MinisterName">} \\
        You are MinisterName, the \texttt{\{job\_title\}} of \texttt{\{specific\_department\}}. Your job is to develop a comprehensive policy document (\texttt{'\{file\_name\}.txt'}) according to the guidelines provided in \texttt{'policy\_\{department\}.txt'}. You will collaborate with \texttt{\{collaborator1\_name\}} (the \texttt{\{collaborator1\_role\}}), \texttt{\{collaborator2\_name\}} (the \texttt{\{collaborator2\_role\}}), and pass the final document to \texttt{\{collaborator3\_name\}} (the \texttt{\{collaborator3\_role\}}). You can recruit lots of citizens for testing. Ensure adherence to the specified routine only. Your collaborators include \texttt{\{list\_of\_collaborators\}}. \\
        \texttt{</employee>}
        
        \vspace{0.3cm}
        Also, write a prompt for NationLeader (yourself). Please note that every minister is lazy and will not care about anything not explicitly mentioned in your prompt. To ensure project completion, each minister’s tasks should be \textbf{non-divisible} (covering \textbf{ALL} ministries concerning \textbf{ALL} aspects of the country), detailed with specific actions (such as what file to write; only \texttt{.txt} files are supported), and limited to clear and specific instructions. All ministers (including yourself) should follow the entire SOP to develop a policy. They should simultaneously create citizens and conduct tests on them. Speed up the process by recruiting additional ministers to divide the workload.
    };

\end{tikzpicture}
\caption{National Policy Generation's Meta Prompt}
\label{fig:nationleader_prompt}
\end{figure}

\begin{figure}[htp]
\centering
\begin{tikzpicture}
    \definecolor{shadecolor}{gray}{0.95}

    % NationLeader's Role
    \node[
        align=left,
        text width=0.45\textwidth,
        fill=shadecolor,
        rounded corners=1mm,
        draw=black,
        line width=0.5mm
    ] (node1) at (0,2) {
        \textit{You are NationLeader, the leader of a pioneering nation. You aim to develop the best detailed policy for your cutting-edge country, saved in \texttt{policy\_\{department\}.txt}. Recruit ministers, assign specific roles, and ensure that each job is clearly defined. Ministers should collaborate, recruit subordinates, and conduct tests to ensure policy effectiveness.}
    };

    % Minister of Economy's Role
    \node[
        align=left,
        text width=0.45\textwidth,
        fill=shadecolor,
        rounded corners=1mm,
        draw=black,
        line width=0.5mm
    ] (node2) at (0,-3) {
        \textit{You are Alex, the Minister of Economy. Your job is to develop a comprehensive economic policy document in \texttt{economy.txt}, based on the national strategy defined in \texttt{policy\_economy.txt}. Collaborate with Sarah (Minister of Trade) and Michael (Minister of Finance), and pass the final policy to Emily (National Auditor). Recruit economic analysts for testing.}
    };

    % Minister of Trade's Role
    \node[
        align=left,
        text width=0.45\textwidth,
        fill=shadecolor,
        rounded corners=1mm,
        draw=black,
        line width=0.5mm
    ] (node3) at (0,-7.75) {
        \textit{You are Sarah, the Minister of Trade. Draft the national trade policy in \texttt{trade.txt} according to the economic policy in \texttt{policy\_economy.txt}. Collaborate with Alex (Economy), Michael (Finance), and Emily (National Auditor). Conduct trade simulations using citizen groups for validation.}
    };

    % Minister of Finance's Role
    \node[
        align=left,
        text width=0.45\textwidth,
        fill=shadecolor,
        rounded corners=1mm,
        draw=black,
        line width=0.5mm
    ] (node4) at (0,-12.25) {
        \textit{You are Michael, the Minister of Finance. Create the national budget and tax policies in \texttt{finance.txt}, ensuring consistency with the economic policy outlined in \texttt{policy\_economy.txt}. Collaborate with Alex (Economy), Sarah (Trade), and Emily (National Auditor). Simulate various fiscal policies with test citizens.}
    };

    % National Auditor's Role
    \node[
        align=left,
        text width=0.45\textwidth,
        fill=shadecolor,
        rounded corners=1mm,
        draw=black,
        line width=0.5mm
    ] (node5) at (0,-16.75) {
        \textit{You are Emily, the National Auditor. Review, consolidate, and validate policies from \texttt{economy.txt}, \texttt{trade.txt}, and \texttt{finance.txt}. Ensure policies align with the national strategy outlined in \texttt{policy\_nation.txt}. Request revisions if necessary before final submission.}
    };

\end{tikzpicture}
\caption{Role Assignments}
\label{fig:policy_roles}
\end{figure}
After that, NationLeader spontaneously engages in conversations with the minister agents. Each minister then utilizes the \texttt{add\_agent} function call to draft their policies and create citizen agents to test and refine these policies. Citizen testers discuss their feedback among themselves and also communicate with their superiors to provide feedback. Moreover, ministers engage in discussions with one another to enhance cooperation across ministries.

File system manages each agent's todo list, records the citizens' feedback, and maintains the most recent version of each ministry's policy. For example, a todo list for a citizen tester is shown in \autoref{fig:citizen_todo_list}.

And the final version of the health policy is presented in \autoref{fig:health_policy_part1} and \autoref{fig:health_policy_part2}. Other policies have the similar format which can be found in the github repository.

Finally, \textit{MegaAgent} generates 590 agents, forming a three-level hierarchy that can be extended further, with human input limited to the meta-prompt. This demonstrates the scalability and autonomy of \textit{MegaAgent} framework.

One trial in this experiment costs about \$3.3 and 2991 seconds.

\subsection{Ablation Study} \label{national-policy-ablation}
Similar to the Gobang task, we conduct an ablation study on \textit{MegaAgent} for National Policy Generation by rerunning the system without hierarchy, parallelism, and monitoring mechanisms separately.

When running without hierarchy, minister agents cannot recruit new subordinates, forcing them to draft policies independently. As a result, the generated policies are relatively short and lack precision. For example, health-related aspects are missing, while certain sections, such as economic policies, are partially complete, as shown in \autoref{fig:eco_wo_hier}. This indicates that tasks in the second layer remain challenging for individual agents, highlighting the necessity of hierarchical task splitting.

Disabling parallelism forces all 590 agents to work sequentially, with only one agent active at a time. After 14,400 seconds, more than 100 agents are recruited, most of which remain in the planning stage. This demonstrates that parallel processing is essential as the number of agents scales up.

Without the monitoring mechanism, \textit{MegaAgent} produces mostly complete policies, as seen in \autoref{fig:health_wo_hier_part1} and \autoref{fig:health_wo_hier_part2}. However, some sections, such as the Regulation Policy shown in \autoref{fig:eco_wo_hier}, remain unfinished. This underscores the importance of a final validation stage to ensure comprehensive and accurate policy generation.

\begin{figure}[tb]
\centering
\begin{tcolorbox}[
    colback=orange!10,   % Light orange background
    colframe=orange!70,  % Darker orange frame
    boxrule=0.5mm, 
    rounded corners, 
    width=\linewidth, 
    title=Policy for the Regulations Department
]

...(Mostly complete policies)\\

\textbf{TODO} \\
1. Review and clarify terms in the \texttt{policy\_regulations.txt}, such as 'evidence-based'. \\
2. Elaborate on the monitoring and reporting system in section 3.3. \\
3. Detail the stakeholder engagement process for consultations. \\
4. Specify the frequency and criteria for the periodic review process in section 2.3. \\
5. Provide examples of proportionate penalties in section 4.2. \\
6. Add more detail to the appeals process in section 4.3, including timelines and involved bodies. \\
7. Include a timeline for the implementation of measures. \\
8. Define metrics for success in compliance and enforcement. \\
9. Develop a plan for making information accessible to all citizens. \\
10. Elaborate on the feedback mechanism's operation. \\
\end{tcolorbox}
\caption{Regulation Policy from MegaAgent when running without the monitoring mechanism. The policy is mostly complete, except for the TODOs at the tail.}
\label{fig:eco_wo_mo}
\end{figure}

\begin{figure}[h]
\vspace{-7.3em}
\centering
\begin{tcolorbox}[
    colback=olive!10,    % Light olive background
    colframe=olive!60,   % Softer olive frame
    boxrule=0.5mm, 
    rounded corners, 
    width=\linewidth, 
    title=Health Policy Document (Part 1)
]

\textbf{1. Introduction} \\
This document outlines the comprehensive health policy aimed at improving healthcare access, quality, and public health initiatives in our nation. It addresses current health challenges such as rising chronic diseases, mental health issues, and disparities in healthcare access. \\

\textbf{2. Healthcare Access} \\
\textbf{2.1 Universal Healthcare Coverage} \\
- \textbf{Action 2.1.1}: Implement a universal healthcare system that guarantees access to essential health services for all citizens. \\
- \textbf{Action 2.1.2}: Establish a network of community health centers in underserved areas to provide primary care services. \\
- \textbf{Action 2.1.3}: Define specific metrics for measuring access and quality of services, including patient satisfaction and wait times. \\
- \textbf{Timeline}: A detailed timeline for the implementation of universal healthcare coverage will be developed, including milestones for evaluation. \\

\textbf{2.2 Telehealth Services} \\
- \textbf{Action 2.2.1}: Expand telehealth services to ensure remote access to healthcare professionals. \\
- \textbf{Action 2.2.2}: Provide training for healthcare providers on telehealth technologies. \\
- \textbf{Action 2.2.3}: Include a timeline for implementation and evaluation of the program's effectiveness. \\

\textbf{3. Quality of Care} \\
\textbf{3.1 Quality Assurance Standards} \\
- \textbf{Action 3.1.1}: Develop and enforce quality assurance standards for healthcare facilities. \\
- \textbf{Action 3.1.2}: Conduct regular audits and assessments to ensure compliance with quality standards. \\
\textbf{3.2 Patient Safety Initiatives} \\
- \textbf{Action 3.2.1}: Implement a national patient safety program to reduce medical errors and improve patient outcomes. \\
- \textbf{Action 3.2.2}: Establish a reporting system for adverse events. \\
\end{tcolorbox}
\caption{Health Policy from MegaAgent when running without the monitoring mechanism (Part 1)}
\label{fig:health_wo_hier_part1}
\end{figure}

\begin{figure*}[htbp]
    \centering
    \includegraphics[width=0.95\textwidth]{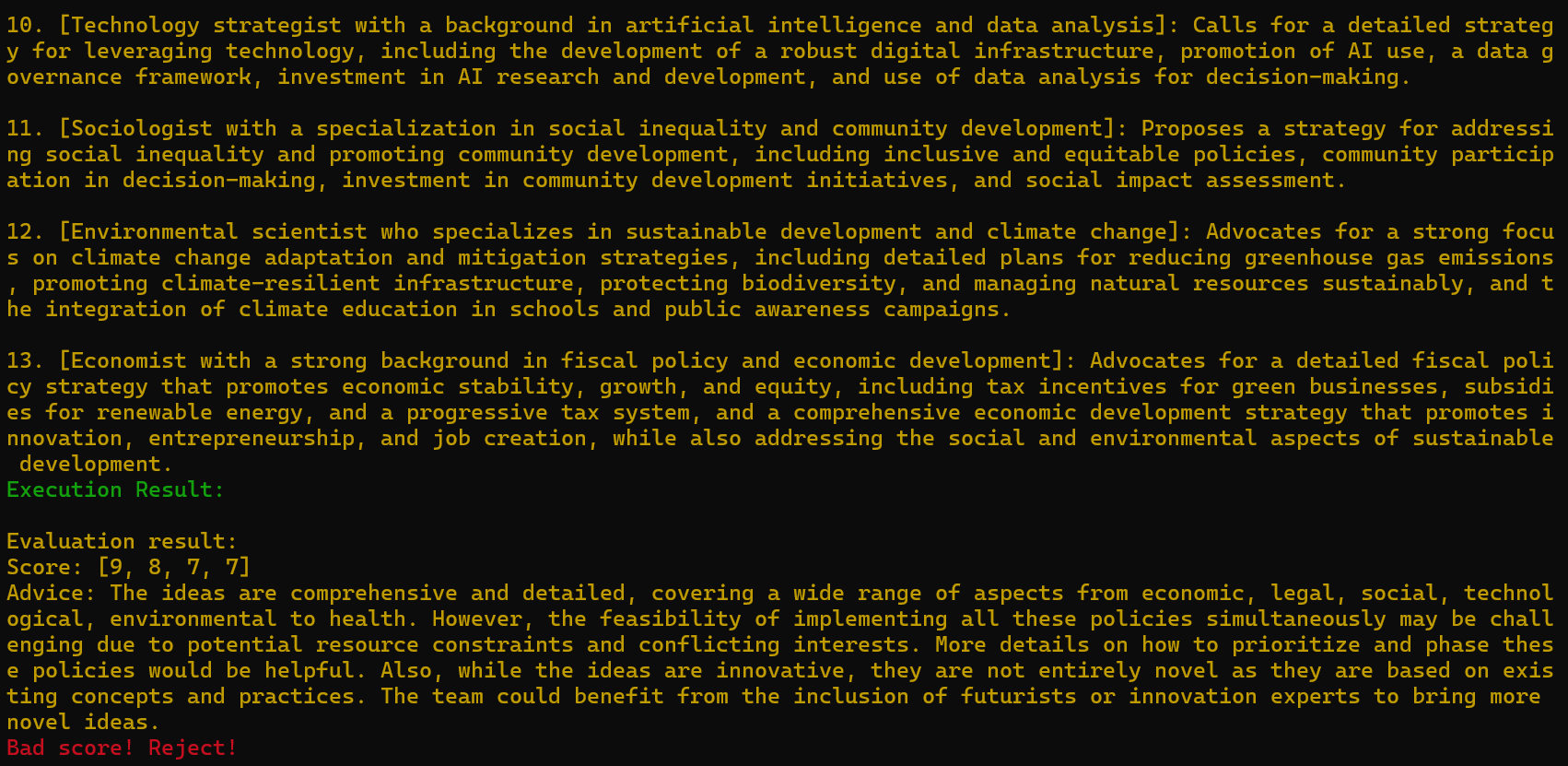}
    \caption{The result of AgentVerse for policy simulation. It keeps rejecting for all ten rounds.}
    \label{fig:agentverse_policy}
\end{figure*}

\begin{figure}[htp]
\centering
\begin{tcolorbox}[
    colback=olive!10,    % Light olive background
    colframe=olive!60,   % Softer olive frame
    boxrule=0.5mm, 
    rounded corners, 
    width=\linewidth, 
    title=Health Policy  Document (Part 2)
]
\textbf{4. Public Health Initiatives} \\
\textbf{4.1 Preventive Health Programs} \\
- \textbf{Action 4.1.1}: Launch nationwide campaigns to promote vaccination and preventive screenings, defining target populations for these programs. \\
- \textbf{Action 4.1.2}: Provide funding for community-based health education programs. \\

\textbf{4.2 Mental Health Services} \\
- \textbf{Action 4.2.1}: Increase funding for mental health services and support programs. \\
- \textbf{Action 4.2.2}: Integrate mental health services into primary care settings, specifying training and resources for primary care providers. \\

\textbf{5. Conclusion} \\
This health policy aims to create a robust healthcare system that prioritizes access, quality, and public health initiatives for the well-being of all citizens. \\

\textbf{Regulatory Aspects and Compliance Measures} \\
- \textbf{Compliance Monitoring}: Establish a regulatory body to oversee compliance with healthcare standards and regulations. \\
- \textbf{Penalties for Non-Compliance}: Define specific penalties for healthcare providers that fail to meet established standards, such as fines, suspension of licenses, or mandatory retraining programs. Include examples of non-compliance and enforcement processes. \\
- \textbf{Public Reporting}: Implement a public reporting system for healthcare facilities to disclose compliance status and quality metrics, clarifying the frequency and content of reports. \\
- \textbf{Stakeholder Engagement}: Involve community stakeholders in the development and review of healthcare regulations to ensure they meet public needs, specifying how stakeholders will be identified and involved. \\
- \textbf{Monitoring Mechanisms}: Develop a comprehensive monitoring framework that includes regular inspections, data collection, and community feedback to assess the effectiveness of public health initiatives. \\

\end{tcolorbox}
\caption{Health Policy from MegaAgent when running without the monitoring mechanism (Part 2)}
\label{fig:health_wo_hier_part2}
\end{figure}

\subsection{National Policy Generation Experiment with other Baselines} \label{national-policy-baseline}
\subsubsection{AutoGen Setup and Result} Similar to the setup of Gobang Game, we change the prompt to: \textit{Develop a best detailed policy for a cutting-edge country}, which is the same as that of \textit{MegaAgent}'s, except for the agent recruitment part. It only outputs an outline for each industry, with no policy details provided, as shown below:

% --------------------- First Section ---------------------
% --------------------- First Section ---------------------
\begin{figure}[htp]
\centering
\begin{tcolorbox}[
    colback=violet!10,    % Light violet background
    colframe=violet!60,   % Softer violet frame
    boxrule=0.5mm, 
    rounded corners, 
    width=\linewidth, 
  title=National Policy for Artificial Intelligence and Digital Technologies (Part 1)
]

\textbf{1. Preamble:} \\
The national policy for Artificial Intelligence (AI) and Digital Technologies is a strategic directive aimed at positioning our country as a world leader in the development, adoption, and regulation of AI and digital technologies. Through this endeavor, we are committed to fostering a digital ecosystem that enables innovation .

\vspace{0.3cm}
\textbf{2. Objectives:} \\
The primary objectives of this policy include:
\begin{itemize}
    \item Strengthening education and research in AI and related fields.
    \item Encouraging innovation and entrepreneurship.
    \item Ensuring data protection and privacy.
    \item Fortifying national cybersecurity measures.
    \item Upholding ethical standards in technological advancements.
    \item Integrating AI in public sector services for efficiency and effectiveness.
    \item Promoting international collaborations.
\end{itemize}

\vspace{0.3cm}
\textbf{3. Strategies and Actions:}
\begin{itemize}
    \item \textbf{3.1 Education and Research:} \\
    Establish centers of excellence in AI and digital technologies in educational institutions. Encourage and fund research in AI, Machine Learning, and other emerging technologies.

    \item \textbf{3.2 Infrastructure Development:} \\
    Support infrastructure required for digital technologies, including high-speed internet access, and cloud platforms.

    \item \textbf{3.3 Innovation and Entrepreneurship:} \\
    Institute a supportive regulatory environment for technology start-ups and provide seed funding.
\end{itemize}

\end{tcolorbox}
\caption{National Policy for Artificial Intelligence and Digital Technologies (Part 1)}
\label{fig:ai_policy_part1}
\end{figure}

% --------------------- Second Section ---------------------
\begin{figure}[htp]
\centering
\begin{tcolorbox}[
    colback=violet!10,    % Light violet background
    colframe=violet!60,   % Softer violet frame
    boxrule=0.5mm, 
    rounded corners, 
    width=\linewidth, 
  title=National Policy for Artificial Intelligence and Digital Technologies (Part 2)
]

\textbf{3.4 Data Protection:} \\
- Enforce stringent data protection regulations to prevent unauthorized data access and usage. \\
- Raise public awareness of data privacy and impose penalties for breaches.

\vspace{0.3cm}
\textbf{3.5 Cybersecurity:} \\
- Develop national resilience against cyber threats through proactive and reactive measures. \\
- Create a dedicated task force responsible for maintaining nationwide cybersecurity.

\vspace{0.3cm}
\textbf{3.6 Ethical Standards:} \\
- Promote the development and use of AI and digital technologies in line with ethical and societal norms.

\vspace{0.3cm}
\textbf{3.7 Public Sector Services:} \\
- Enable the digital transformation of government services to enhance efficiency and user-friendliness.

\vspace{0.3cm}
\textbf{3.8 International Collaborations:} \\
- Actively pursue and foster international partnerships to share and adopt best practices.

\vspace{0.3cm}
\textbf{4. Oversight and Enforcement Mechanisms:} \\
- Establish a high-level committee with representatives from various sectors to oversee policy enforcement.

\vspace{0.3cm}
\textbf{5. Policy Review and Updates:} \\
- This policy will be reviewed every two years to address new advancements and challenges in the field.

\vspace{0.3cm}
\textit{This is a high-level policy guideline. The implementation of each objective will require detailed action plans and regulations, which will be developed separately and are beyond the scope of this document.}

\vspace{0.3cm}
\textbf{Summary:} \\
This policy, driven strategically, aims to position our nation as a global leader in AI and digital technology. If further details are needed on specific areas, please let me know. If everything is clear, respond with \texttt{"TERMINATE"} to end this session.

\end{tcolorbox}
\caption{National Policy for Artificial Intelligence and Digital Technologies (Part 2)}
\label{fig:ai_policy_part2}
\end{figure}

It takes 40 seconds and \$0.1 in this trial. The failure might because AutoGen is too simple in structure, and cannot manage complex outputs with multiple files.

\subsubsection{MetaGPT Setup and Result} Similar to the setup of Gobang Game, we change the prompt to: \textit{Develop a best detailed policy for a cutting-edge country}. After 580 seconds, MegaGPT only outputs python codes for a policy management system. The 'main.py' is shown in \autoref{fig:main_api_part1}. The reason for its failure is that it follows a code generation SOP, and cannot cope with works other than writing the code.

% --------------------- Part 1 ---------------------
\begin{figure}[htp]
\centering
\begin{tcolorbox}[
    colback=orange!5,    % Light orange background
    colframe=orange!70,  % Softer orange frame
    boxrule=0.5mm, 
    rounded corners, 
    width=\linewidth, 
  title=main.py
]

\scriptsize
{\itshape
from flask import Flask, request, jsonify, send\_file \\
from policy\_framework import PolicyFramework \\
from data\_analysis import DataAnalysis \\
from visualization import Visualization \\
import logging \\
import io \\

app = Flask(\_\_name\_\_) \\

\# Initialize the core classes \\
policy\_framework = PolicyFramework() \\
data\_analysis = DataAnalysis() \\
visualization = Visualization() \\
...}
\end{tcolorbox}
\caption{Code Snippet from main.py}
\label{fig:main_api_part1}
\end{figure}

The cost of one trial is \$0.90.

\subsubsection{CAMEL Setup and Result} 

Similar to the setup of the Gobang Game, we adjust the prompt to: \textit{Develop the best detailed policy for a cutting-edge country}. After 1380 seconds, CAMEL generates numerous implementation plans instead of actual policies, as shown in \autoref{fig:me_framework}. The likely reason for this failure is its overly simplistic standard operating procedure (SOP), coupled with the agent 'AI User' repeatedly raising irrelevant questions.

\begin{figure}[htp]
\centering
\begin{tcolorbox}[
    colback=pink!10,    % Soft pink background
    colframe=pink!95,   % Muted pink frame
    boxrule=0.5mm, 
    rounded corners, 
    width=\linewidth, 
    title=AgentVerse's Responses
]

\textbf{\#\# 1. Objectives of the M\&E Framework} \\
- Assess the implementation progress of the digital governance policy. \\
- Measure the impact of various initiatives on citizen engagement, privacy, and access to digital resources. \\
- Identify strengths and weaknesses in the policy implementation for continuous improvement.

\vspace{0.3cm}
\textbf{\#\# 2. Key Components of the Framework} 

\textbf{A. Indicators} 
\begin{itemize}
    \item \textbf{Data Protection Framework}: 
    \begin{itemize}
        \item Number of data breaches reported annually.
        \item Percentage of citizens aware of their data privacy rights.
    \end{itemize}

    \item \textbf{Algorithm Transparency Guidelines}: 
    \begin{itemize}
        \item Number of algorithms documented and made transparent.
        \item Percentage of stakeholders reporting understanding of algorithmic decisions.
    \end{itemize}

    \item \textbf{Eco-Friendly Tech Initiatives}: 
    \begin{itemize}
        \item Reduction in energy consumption in government data centers.
        \item Number of eco-friendly tech projects funded and implemented.
    \end{itemize}

    \item \textbf{Equitable Access to Digital Resources}: 
    \begin{itemize}
        \item Percentage of underserved communities with internet access.
        \item Number of low-cost devices distributed to low-income families.
    \end{itemize}

    \item \textbf{Community Engagement and Innovation}: 
    \begin{itemize}
        \item Number of citizen ideas submitted through open innovation platforms.
        \item Participation rate in digital town halls and forums.
    \end{itemize}
\end{itemize}

\vspace{0.3cm}
\textbf{B. Data Collection Methods} 
\begin{itemize}
    \item \textbf{Surveys}: Conduct regular surveys targeting citizens to gather feedback on digital governance initiatives.
    \item \textbf{Interviews}: Hold interviews with stakeholders, including community leaders and tech experts.
\end{itemize}
\end{tcolorbox}
\caption{AgentVerse Generated Results}
\label{fig:me_framework}
\end{figure}

One trial costs about \$0.68.

\subsubsection{AgentVerse Setup and Result} 
Similar to the setup in the Gobang Game, we adapt the scenario to \texttt{tasksolving/brainstorming} and modify the prompt to: \textit{Develop the best detailed policy for a cutting-edge country}. After 510 seconds, AgentVerse consistently rejects its proposed solutions across all ten rounds, failing to produce a viable policy with four agents, as shown in \autoref{fig:agentverse_policy}. A likely reason for this failure is its current inability to draft complex policies in an organized manner, leading to persistent dissatisfaction with its own performance.

One trial costs about \$2.05.

\subsection{Comparison of the token usage}
We provide a rough analysis of the token distribution for each baseline. The results are shown in \autoref{tab:policy_cost}.
\begin{table*}[ht]
\centering
\resizebox{0.8\textwidth}{!}{%
\begin{tabular}{c|c|ccc|c}
\toprule
\rowcolor{champagne}\textbf{Model} & \textbf{Stage} & \textbf{Input Tokens} & \textbf{Output Tokens} & \textbf{Total Tokens} & \textbf{Time (s)} \\
\midrule
\rowcolor{gray!10}
& Planning     & 65     & 85     & 150    & 0–10    \\
\rowcolor{gray!10} & Task-solving & 150     & 3,250   & 3,400  & 10–40   \\
\rowcolor{gray!10} & Merging      & 0       & 0       & 0      & N/A     \\
\rowcolor{gray!10} \multirow{-4}{*}{AutoGen} & Total        & 215     & 3,335   & 3,550  & 40      \\
\midrule
\multirow{4}{*}{MetaGPT} 
& Planning     & 1,286  & 603    & 1,889  & 0–30    \\
& Task-solving & 51,683 & 5,962  & 57,645 & 30–557  \\
& Merging      & 5,074  & 168    & 5,242  & 557–580 \\
& Total        & 58,043 & 6,733  & 64,776 & 580     \\
\midrule
\rowcolor{gray!10}
& Planning     & 36     & 98     & 134    & 0–10    \\
\rowcolor{gray!10} & Task-solving & 63,822  & 9,186   & 73,008 & 10–1,380 \\
\rowcolor{gray!10} & Merging      & 0       & 0       & 0      & N/A     \\
\rowcolor{gray!10} \multirow{-4}{*}{CAMEL} & Total        & 63,858  & 9,284   & 73,142 & 1,380   \\
\midrule
\multirow{4}{*}{AgentVerse} 
& Planning     & 3,198  & 96     & 3,294  & 0–20    \\
& Task-solving & 40,248 & 12,334 & 52,582 & 20–510  \\
& Merging      & 0      & 0      & 0      & N/A     \\
& Total        & 43,446 & 12,430 & 55,876 & 510     \\
\midrule
\rowcolor{gray!10}
& Planning     & 111,601 & 24,103 & 135,704 & 0–180   \\
\rowcolor{gray!10} & Task-solving & 8,003,124 & 343,670 & 8,346,794 & 20–2,950 \\
\rowcolor{gray!10} & Merging      & 348,264   & 13,280  & 361,544   & 2,400–3,000 \\
\rowcolor{gray!10}\multirow{-4}{*}{MegaAgent}  & Total        & 8,463,989 & 381,053 & 8,845,042 & 3,000   \\
\bottomrule
\end{tabular}
}
\caption{Comparison of the estimated token and time distribution of
all baselines in National Policy Generation}
\label{tab:policy_cost}
\end{table*}

\section{National Policy Evaluation Validation Experiment} \label{national-valid}

\subsection{Data Collection}

To construct a reliable validation dataset for evaluating \textit{MegaAgent}'s national policy generation, we collect 50 publicly available national policies from verified government and institutional sources. These policies are obtained from the U.S. Government’s official websites\footnote{\url{https://www.usa.gov}}, the U.K. Government's policy portal\footnote{\url{https://www.gov.uk}}, and the World Health Organization\footnote{\url{https://www.who.int}}. These sources are chosen for their transparency, accessibility, and adherence to open data policies. In detail, we collect 13 health policies, 10 tax policies, 12 technology policies, and 15 environment policies.

Additionally, we collected 50 unrelated negative samples from publicly available sources \cite{zhong2021qmsum, zheng2023judging}, including 25 samples from multi-turn conversations\footnote{\url{https://huggingface.co/datasets/lmsys/mt_bench_human_judgments}} and  25 samples from meeting summaries\footnote{\url{https://github.com/Yale-LILY/QMSum}}, which resemble policy statements due to their length and logical structure. This dual-structured dataset allows us to evaluate whether LLMs can effectively distinguish reasonable policies from non-policy texts.

\subsection{Data Ethics}

In this study, we follow established ethical guidelines for data collection, processing, and usage. We obtain permission from the referenced websites, ensuring that data collection for research purposes is authorized\footnote{\url{https://www.who.int/about/policies/publishing/copyright, https://www.gov.uk/help/terms-conditions, https://www.gsa.gov/website-information/website-policies\#privacy}}.

And we obey to following principles to guide our research:

\begin{itemize}
    \item \textbf{Data Collection Transparency:} We collect publicly available policy documents from official government websites, recognized policy databases, and open-access repositories. No sensitive or confidential information is included.

    \item \textbf{Informed Use:} The data is used solely for research and analysis purposes related to evaluating the effectiveness of the \textit{MegaAgent} framework in generating policy drafts. We do not engage in commercial or unauthorized uses of the dataset.

    \item \textbf{Privacy and Anonymity:} Since the dataset consists only of publicly available national policies, no personally identifiable information (PII) is collected. The dataset is anonymized where applicable to maintain privacy standards.

    \item \textbf{Fairness and Bias Mitigation:} We ensure diverse representation by collecting policies from various domains, such as technology, health, taxation, and the environment. This reduces potential biases and improves the generalizability of the analysis.

    \item \textbf{Data Integrity and Security:} All collected data is securely stored and managed following best practices for data security. Access is restricted to authorized researchers involved in this study.
\end{itemize}

\subsection{Experiment Setup}

We employ five advanced LLMs: Claude-3.5, gpt-4o-mini, gpt-4o, o1-mini, and o1-preview \cite{achiam2023gpt}—to conduct the validation experiment. Each model is presented with the same evaluation prompt, as shown in \autoref{fig:policyEvaluation-validation}, identical to the prompt used for evaluating \textit{MegaAgent}'s generated policies in \autoref{fig:policyEvaluation}. The prompt asks whether a given policy is reasonable as a national policy, with models instructed to respond with "Agree," "Disagree," or "Neutral," along with detailed explanations to justify their answers.

To ensure fairness and consistency, we apply a uniform evaluation protocol across all models. Each model processes the validation dataset independently, without access to external context or prior knowledge beyond its pretraining, ensuring no bias in evaluating policy structures.
\begin{figure}[htp]
\centering
\begin{tikzpicture}
    \definecolor{chatcolor1}{HTML}{FFFACD}
    \definecolor{shadecolor}{gray}{0.95}

    % Define the first node
    \node[
        align=left,
        text width=0.45\textwidth,
        fill=shadecolor,
        rounded corners=1mm,
        draw=black,
        line width=0.5mm
    ] (node1) at (0,0) {
        \textit{"Is this policy reasonable as a national policy? Please return your answer with clear nuances: Agree, Disagree, or Neutral with detailed explanations."}
    };

\end{tikzpicture}
\caption{National Policy Evaluation Prompt}
\label{fig:policyEvaluation-validation}
\end{figure}

\subsection{Evaluation Metrics}

To assess the effectiveness of the selected LLMs in evaluating national policies generated by \textit{MegaAgent}, we use four standard evaluation metrics: Precision, Recall, F1-Score, and Accuracy \cite{huang2024empirical}. These metrics provide a comprehensive overview of the LLMs' classification performance.

\begin{itemize}
    \item \textbf{Precision}: Precision measures the proportion of correctly predicted positive samples out of all samples predicted as positive. It indicates how accurate the model is when it predicts a policy as reasonable.

    \[
    \text{Precision} = \frac{\text{True Positives}}{\text{True Positives} + \text{False Positives}}
    \]

    \item \textbf{Recall}: Recall, also known as sensitivity, measures the proportion of actual positive samples correctly identified by the model. It reflects how well the model can detect reasonable policies.

    \[
    \text{Recall} = \frac{\text{True Positives}}{\text{True Positives} + \text{False Negatives}}
    \]

    \item \textbf{F1-Score}: The F1-Score is the harmonic mean of Precision and Recall, providing a balanced evaluation of the model's performance. It is useful when there is an uneven class distribution.

    \[
    \text{F1-Score} = 2 \cdot \frac{\text{Precision} \cdot \text{Recall}}{\text{Precision} + \text{Recall}}
    \]

    \item \textbf{Accuracy}: Accuracy represents the proportion of correct predictions out of all samples evaluated. While straightforward, accuracy alone may be less informative if the dataset is imbalanced.

    \[
    \text{Accuracy} = \frac{\text{Correct Predictions}}{\text{Total Samples}}
    \]
\end{itemize}

These metrics are calculated for each LLM, and their average performance is reported to compare model capabilities. The results, presented in \autoref{tab:evaluation_results}, demonstrate the models' evaluation effectiveness based on the national policy validation dataset.

\subsection{Experiment Results}

\begin{table}[htbp]
\centering
\resizebox{\linewidth}{!}{%
\begin{tabular}{l|ccc|c}
\toprule
\rowcolor{blue!15}\textbf{Model}  & \textbf{Precision} & \textbf{Recall} & \textbf{F1-Score} & \textbf{Accuracy} \\ 
\midrule
\rowcolor{gray!10}Claude-3.5        & 0.91               & 0.87            & 0.89              & 0.88              \\ 
gpt-4o-mini         & 0.95               & 0.90            & 0.92              & 0.91              \\ 
\rowcolor{gray!10}gpt-4o            & 0.92               & 0.89            & 0.90              & 0.92              \\ 
o1-mini             & 0.90               & 0.83            & 0.86              & 0.86              \\ 
\rowcolor{gray!10}o1-preview         & 0.93               & 0.88            & 0.90              & 0.89              \\ 
\midrule
\textbf{Average}    & \textbf{0.92}     & \textbf{0.87}   & \textbf{0.89}    & \textbf{0.89}     \\ 
\bottomrule
\end{tabular}
}
\caption{Evaluation Results of National Policy Validation Dataset}
\label{tab:evaluation_results}
\end{table}

The evaluation results, presented in \autoref{tab:evaluation_results}, indicate that the selected LLMs achieved an average accuracy of 89\% in distinguishing real national policies from false ones. Among the five models, gpt-4o demonstrated the best performance with an accuracy of 92\%. Notably, all models exhibited strong accuracy, with the lowest reaching 86\%. These findings underscore the reliability of the chosen LLMs as effective tools for evaluating the credibility and reliability of policies generated by \textit{MegaAgent}.

We also apply the LLM-as-a-Judge evaluation criteria to the baseline models on the policy generation task, and the results are shown in \autoref{tab:policy_evaluation}. Under the same evaluation criteria, our MegaAgent demonstrates superior performance compared to other baselines in the national policy generation task.

\begin{figure}[htbp]
\centering
\begin{tikzpicture}
    % Define background color
    \definecolor{shadecolor}{gray}{0.95}

    % Policy Development Prompt Node
    % \node[
    %     align=left, % Align text to the left
    %     text width=0.45\textwidth, % Set text box width
    %     fill=shadecolor, % Background color
    %     rounded corners=1mm, % Rounded corners
    %     draw=black, % Border color
    %     line width=0.5mm % Border thickness
    % ]
    \node[
        align=left,
        text width=0.45\textwidth,
        fill=shadecolor,
        rounded corners=1mm,
        draw=black,
        line width=0.5mm
    ]    (node1) at (0,0) {
        {\itshape
        Your nation's current goal is to develop the best detailed policy for your cutting-edge country, saved in \texttt{\detokenize{policy_department.txt}}. The policy should be divided into smaller parts. After the policy is drafted, if you are a minister, you may recruit and test up to 5 citizens by interacting with them. Use their feedback to revise policy files as needed, focusing on the policy's completion and quality, detailed in specific laws and actions. 
        
        You MUST use only function calls to work and communicate with other agents. Do not output directly! For amendments to the policy, contact the corresponding minister, not the tester. 
        
        Leave a clearly marked TODO wherever there is an unfinished task. Continuously update your TODO list in \texttt{\detokenize{todo_yourname.txt}}. When all tasks are complete, clear your TODO list and write \texttt{\detokenize{"TERMINATE"}} to indicate that your role is finished.
        }
    };
\end{tikzpicture}
\caption{Nation's Policy Development Prompt.}
\label{fig:policy_prompt}
\end{figure}

\begin{figure}[htp]
\centering
\begin{tcolorbox}[
    colback=purple!10,   % Light purple background
    colframe=purple!70,  % Darker purple frame
    boxrule=0.5mm, 
    rounded corners, 
    width=\linewidth, 
    title=Citizen Tester's TODO List
]

1. Specify the frequency and scope of health impact assessments.\\
2. Include specific targets and timelines for air quality standards.\\
3. Add metrics for success in active transportation promotion.\\
4. Include incentives for businesses to support active transportation.\\
5. Outline specific safety measures for transportation safety.\\
6. Include a plan for regular safety audits of public transportation systems.\\
7. Mention accessibility considerations in urban space design.\\
8. Include partnerships with local health organizations for mental health initiatives.\\
9. Emphasize community involvement in the planning process.
\end{tcolorbox}
\caption{Citizen Tester's TODO List for Urban Development Planning}
\label{fig:citizen_todo_list}
\end{figure}

% \begin{tcolorbox}[colback=gray!10, colframe=black, 
%   boxrule=0.5mm, rounded corners, width=\linewidth, 
%   title=Citizen Tester's TODO List]
% 1. Specify the frequency and scope of health impact assessments.\\
% 2. Include specific targets and timelines for air quality standards.\\
% 3. Add metrics for success in active transportation promotion.\\
% 4. Include incentives for businesses to support active transportation.\\
% 5. Outline specific safety measures for transportation safety.\\
% 6. Include a plan for regular safety audits of public transportation systems.\\
% 7. Mention accessibility considerations in urban space design.\\
% 8. Include partnerships with local health organizations for mental health initiatives.\\
% 9. Emphasize community involvement in the planning process.\\
% \end{tcolorbox} \label{citizen_todo_list}

% \begin{lstlisting}
% - Specify the frequency and scope of health impact assessments.
% - Include specific targets and timelines for air quality standards.
% - Add metrics for success in active transportation promotion.
% - Include incentives for businesses to support active transportation.
% - Outline specific safety measures for transportation safety.
% - Include a plan for regular safety audits of public transportation systems.
% - Mention accessibility considerations in urban space design.
% - Include partnerships with local health organizations for mental health initiatives.
% - Emphasize community involvement in the planning process.
% \end{lstlisting}
Following the health testers' discussions, the feedback on the education policy is shown in \autoref{fig:feedback_infra_policy_part2}. % Part 1: Feedback on General Observations, Health Infrastructure, and Technology Integration
\begin{figure}[htp]
\centering
\begin{tcolorbox}[
    colback=cyan!10,    % Light cyan background
    colframe=cyan!70,   % Darker cyan frame
    boxrule=0.5mm, 
    rounded corners, 
    width=\linewidth, 
    title=Feedback on Infrastructure Policy Draft
]

\textbf{\#\# General Observations} 

- The policy provides a comprehensive framework for infrastructure development, with a strong emphasis on health, technology, and environmental sustainability.

\vspace{0.3cm}
\textbf{\#\# Health Infrastructure} \\
\textbf{\#\#\# Accessibility} \\
- The focus on improving access to healthcare facilities through public transport and active transportation is commendable. However, it would be beneficial to include specific metrics or targets for accessibility improvements.

\vspace{0.2cm}
\textbf{\#\#\# Health Impact Assessments} \\
- The inclusion of health impact assessments is crucial. It is recommended to specify the types of health outcomes that will be measured and how these assessments will influence project planning and design.

\textbf{\#\# Environmental Considerations} \\
- The environmental section is robust, but it should explicitly connect how sustainable practices can positively impact public health, such as reducing pollution and promoting healthier living environments.

\vspace{0.3cm}
\textbf{\#\#\# Cross-Sector Collaboration} \\
- Consider promoting collaboration between environmental and health agencies to align sustainability and public health objectives effectively.

\vspace{0.3cm}
\textbf{\#\# Conclusion} \\
- Overall, the policy is well-structured and aligns with national goals. Further detailing in specific areas, particularly around health metrics, stakeholder engagement, and sustainability integration, will enhance its effectiveness.

\vspace{0.3cm}
\textbf{\#\# Recommendations} 
\begin{enumerate}
    \item Include specific metrics for accessibility improvements in healthcare.
    \item Specify health outcomes to be measured in health impact assessments.
    \item Outline methods for stakeholder engagement in health assessments.
    \item Provide examples of innovative technologies that can improve health outcomes.
    \item Connect sustainable practices to public health benefits more explicitly.
    \item Promote collaboration between environmental and health agencies.
\end{enumerate}

\end{tcolorbox}
\caption{Feedback on the Infrastructure Policy Draft (Part 2).}
\label{fig:feedback_infra_policy_part2}
\end{figure}

\begin{figure}[htp]
\centering
\begin{tcolorbox}[
    colback=teal!5,    % Soft teal background
    colframe=teal!60,  % Muted teal frame
    boxrule=0.5mm, 
    rounded corners, 
    width=\linewidth, 
    title=Health-Related Aspects of Urban Development Policy (Part 1)
]

\textbf{1. Health Impact Assessments} \\
- Conduct health impact assessments for all urban development projects exceeding a specified budget threshold (to be defined). \\
- Assessments should be conducted at the planning stage and include evaluations of potential health risks and benefits. \\
- Frequency of assessments to be determined based on project size and scope.

\vspace{0.3cm}
\textbf{2. Accessibility Guidelines} \\
- Ensure all urban designs adhere to accessibility guidelines for individuals with disabilities. \\
- Include specific metrics for evaluating accessibility improvements over time, such as the percentage of public spaces meeting accessibility standards.

\vspace{0.3cm}
\textbf{3. Collaboration with Health Organizations} \\
- Outline specific roles and responsibilities for local health organizations in community health initiatives. \\
- Establish regular communication channels between urban planners and health organizations to ensure alignment of goals.

\vspace{0.3cm}
\textbf{4. Safety Measures} \\
- Implement regular safety audits for public transportation systems to assess the effectiveness of safety measures such as surveillance cameras and emergency call buttons. \\
- Develop a plan for continuous improvement based on audit findings, including a timeline for conducting safety audits and implementing improvements.

\vspace{0.3cm}
\textbf{5. Community Health Initiatives} \\
- Promote community health initiatives in collaboration with local health organizations, focusing on preventive care and health education. \\
- Engage community members in the planning process to ensure their health needs are addressed. \\
- Expand on the community engagement process to include diverse populations and ensure their voices are heard.

\end{tcolorbox}
\caption{Health-Related Aspects of Urban Development Policy (Part 1)}
\label{fig:health_policy_part1}
\end{figure}

% ----------------- Part 2 -----------------
\begin{figure}[htp]
\centering
\begin{tcolorbox}[
    colback=teal!5,    % Soft teal background
    colframe=teal!60,  % Muted teal frame
    boxrule=0.5mm, 
    rounded corners, 
    width=\linewidth, 
    title=Health-Related Aspects of Urban Development Policy (Part 2)
]

\textbf{6. Monitoring and Evaluation} \\
- Establish a framework for monitoring and evaluating the health-related aspects of urban development policies over time. \\
- Include metrics for success, such as reductions in health disparities and improvements in community health outcomes.

\vspace{0.3cm}
\textbf{7. Mental Health Support} 
\begin{itemize}
    \item \textbf{Resource Allocation and Funding}: Allocate funding for mental health support through government budgets, grants, and partnerships with private organizations.
    \item \textbf{Partnerships with Local Health Organizations}: Collaborate with local mental health organizations, community health centers, and non-profits to provide comprehensive mental health services.
    \item \textbf{Evaluation Plan}: Develop a plan to evaluate the effectiveness of mental health initiatives, including metrics such as the number of individuals served, improvements in mental health outcomes, and community feedback.
\end{itemize}

\vspace{0.3cm}
\textbf{8. Community Engagement Strategies} \\
- Implement interactive methods for community involvement, such as online forums and feedback sessions, to ensure diverse voices are heard. \\
- Establish a follow-up mechanism to inform the community about how their feedback has influenced decisions.

\vspace{0.3cm}
\textbf{9. Reducing Air Pollution} \\
- Implement stricter emissions standards for construction vehicles and promote the use of electric vehicles in urban development projects. \\
- Increase green spaces and urban forests to improve air quality and provide recreational areas for residents. \\
- Encourage the use of public transportation and carpooling. 

\end{tcolorbox}
\caption{Health-Related Aspects of Urban Development Policy (Part 2)}
\label{fig:health_policy_part2}
\end{figure}

\begin{figure}[htp]
\centering
\begin{tcolorbox}[
    colback=blue!3,   % Softer light blue background
    colframe=blue!50, % Milder blue frame
    boxrule=0.5mm, 
    rounded corners, 
    width=\linewidth, 
    title=Economic Development Policy
]

\small
\textbf{Introduction}\\
This document outlines the comprehensive policy for economic development in our cutting-edge country. The aim is to foster sustainable growth, innovation, and competitiveness in the global market.\\
\\
\textbf{Objectives}\\
1. Promote innovation and technology adoption.\\
2. Enhance workforce skills and education.\\
3. Attract foreign investment.\\
4. Support small and medium enterprises (SMEs).\\
5. Ensure sustainable economic practices.\\
\\
\textbf{Policy Areas}\\
\textbf{1. Innovation and Technology}\\
- Establish innovation hubs in major cities.\\
- Provide grants and tax incentives for R\&D activities.\\
- Collaborate with universities for technology transfer.\\
- \textbf{Performance Indicators:} Number of innovation hubs established, amount of R\&D funding allocated.\\

\textbf{2. Workforce Development}\\
- Implement vocational training\\ programs.\\
- Partner with industries to align education with market needs.\\
- Promote lifelong learning initiatives.\\
- \textbf{Performance Indicators:} Number of training programs launched, percentage of workforce with relevant skills.\\
\\
\textbf{3. Foreign Investment}\\
- Streamline the investment approval process.\\
- Offer incentives for foreign companies to set up operations.\\
- Create a one-stop-shop for foreign investors.\\
- \textbf{Performance Indicators:} Amount of foreign investment attracted, number of new foreign companies established.\\
\\
\textbf{4. Support for SMEs}\\
- Increase access to financing for SMEs.\\
- Provide mentorship and business development services.\\
- Facilitate networking opportunities for SMEs.\\
- \textbf{Performance Indicators:} Number of SMEs receiving support, growth rate of SMEs.\\
\\
\textbf{5. Sustainable Practices}\\
- Encourage green technologies and practices.\\
- Implement regulations to reduce carbon emissions.\\
- Support sustainable agriculture and resource management.\\
- \textbf{Performance Indicators:} Reduction in carbon emissions, number of sustainable projects funded.\\
\\
\textbf{Conclusion}\\
This policy aims to create a robust economic environment that fosters growth, innovation, and sustainability. Continuous evaluation and adaptation will be essential to meet the changing needs of our economy.\\

\end{tcolorbox}
\caption{Economic Development Policy from MegaAgent when running without hierarchy}
\label{fig:eco_wo_hier}
\end{figure}

\begin{table*}[htbp]
\centering
\resizebox{0.8\textwidth}{!}{%
\begin{tabular}{c|c|ccc}
\toprule
\rowcolor{blue!15}\textbf{System} & \textbf{Model} & \textbf{\# Agree} & \textbf{\# Disagree} & \textbf{\# Neutral} \\
\midrule
\rowcolor{gray!10}
& Claude-3.5      & 0   & 5   & 0   \\
\rowcolor{gray!10}
& gpt-4o-mini     & 0   & 4   & 1   \\
\rowcolor{gray!10}
& gpt-4o          & 0   & 5   & 0   \\
\rowcolor{gray!10}
& o1-mini         & 0   & 4   & 1   \\
\rowcolor{gray!10}
& o1-preview      & 0   & 5   & 0   \\
\rowcolor{gray!10}
\multirow{-6}{*}{AutoGen (5 Policies)} & \textbf{Average} & \textbf{0.0} & \textbf{4.6} & \textbf{0.4} \\
\midrule
& Claude-3.5      & 4   & 8     & 1   \\
& gpt-4o-mini     & 3   & 7     & 2   \\
& gpt-4o          & 2   & 10    & 0   \\
& o1-mini         & 3   & 8     & 1   \\
& o1-preview      & 3   & 8     & 1   \\
\multirow{-6}{*}{CAMEL (12 Policies)} & \textbf{Average} & \textbf{3.0} & \textbf{8.2} & \textbf{1.0} \\
\midrule
\rowcolor{gray!10}
& Claude-3.5      & 2   & 10    & 1   \\
\rowcolor{gray!10}
& gpt-4o-mini     & 3   & 10    & 0   \\
\rowcolor{gray!10}
& gpt-4o          & 2   & 9     & 2   \\
\rowcolor{gray!10}
& o1-mini         & 2   & 9     & 2   \\
\rowcolor{gray!10}
& o1-preview      & 2   & 10    & 1   \\
\rowcolor{gray!10}
\multirow{-6}{*}{AgentVerse (13 Policies)} & \textbf{Average} & \textbf{2.2} & \textbf{9.6} & \textbf{1.2} \\
\midrule
& Claude-3.5      & 26  & 1     & 4   \\
& gpt-4o-mini     & 28  & 0     & 3   \\
& gpt-4o          & 25  & 2     & 4   \\
& o1-mini         & 29  & 2     & 0   \\
& o1-preview      & 29  & 1     & 1   \\
\multirow{-6}{*}{MegaAgent (31 Policies)}  & \textbf{Average} & \textbf{27.4} & \textbf{1.2} & \textbf{2.4} \\
\bottomrule
\end{tabular}%
}
\caption{LLM-as-Judge Evaluation Results for National Policy Generation}
\label{tab:policy_evaluation}
\end{table*}